%% file: main.tex
\documentclass[10pt,conference]{IEEEtran}
\usepackage[T1]{fontenc}
\usepackage[utf8]{inputenc}
\usepackage{cite}
\usepackage{amsmath,amssymb,amsfonts}
\usepackage{graphicx}
\usepackage{textcomp}
\usepackage{microtype}
\usepackage{booktabs}
\usepackage{enumitem}
\usepackage{multirow}
\usepackage[table]{xcolor}
\usepackage[most]{tcolorbox}
\usepackage{xurl}
\usepackage{hyperref}

\newif\ifreviewchanges
\reviewchangesfalse
\newcommand{\rev}[1]{{\ifreviewchanges\color{blue}\fi#1}}
\newcommand{\benchname}{\rev{GUIOpsBench}}
\newcommand{\preprintgraphic}[2][]{%
  \ifreviewchanges
    \includegraphics[#1]{fig/review/#2}%
  \else
    \includegraphics[#1]{fig/#2}%
  \fi
}
\hypersetup{
    hidelinks,
    pdftitle={GUIOpsBench: Evaluating Multimodal Agents on GUI-Centric Cloud Operations},
    pdfauthor={Haiyu Huang; Jinyang Liu; Zhihan Jiang; Xiao He; Jiewei Lyu; Yichen Li; Yulun Wu; Tieying Zhang; Michael R Lyu},
    pdfsubject={A benchmark for GUI-based cloud operations},
    pdfkeywords={GUIOpsBench, cloud operations, GUI agents, vision-language models}
}
\title{\rev{GUIOpsBench: Evaluating Multimodal Agents\\
on GUI-Centric Cloud Operations}}
\author{
\IEEEauthorblockN{\rev{Haiyu Huang\IEEEauthorrefmark{1},
Jinyang Liu\IEEEauthorrefmark{2}, Zhihan Jiang\IEEEauthorrefmark{1},
Xiao He\IEEEauthorrefmark{2}, Jiewei Lyu\IEEEauthorrefmark{3},}\\
\rev{Yichen Li\IEEEauthorrefmark{1}, Yulun Wu\IEEEauthorrefmark{1},
Tieying Zhang\IEEEauthorrefmark{2}, Michael R Lyu\IEEEauthorrefmark{1}}}
\IEEEauthorblockA{\rev{
\IEEEauthorrefmark{1}The Chinese University of Hong Kong,
\{hyhuang25, zhjiang22, ycli21, ylwu24, lyu\}@cse.cuhk.edu.hk}\\
\rev{\IEEEauthorrefmark{2}Bytedance,
\{jinyang.liu, xiao.hx, tieying.zhang\}@bytedance.com}\\
\rev{\IEEEauthorrefmark{3}Individual Researcher, jiewei.lyu@gmail.com}}
}

\begin{document}
\maketitle
\begin{abstract}
\rev{Routine cloud administration requires operators to interpret visual dashboards, navigate dense interfaces, and follow procedural instructions across multiple console pages. These workflows are not directly assessed by telemetry-centered AIOps benchmarks or general-purpose GUI-agent benchmarks. Motivated by an industrial study of routine operations, we introduce GUIOpsBench, a benchmark for evaluating vision-language models on cloud management consoles. GUIOpsBench contains 21,680 annotated samples from AWS, Google Cloud, Alibaba Cloud, and Volcengine, organized into Ops-UI Understanding, Ops-Action Grounding, and Ops-Task Completion. A dual-path collection engine combines SOP-guided trajectory generation with exploration-driven semantic reconstruction to capture both documented procedures and a broader range of interface states. Evaluations of seven proprietary and open-source models reveal substantial limitations in domain-specific understanding and precise visual interaction. The best average grounding accuracy is 19.09\%, and the highest average task success rate using screenshots alone is 13.7\%. With DOM-assisted grounding and multi-agent execution in TrajMAS, the highest average task success rate rises to 50.1\%, while substantial room for improvement remains. These results identify visual grounding and operational knowledge as key challenges for reliable GUI-based cloud automation.}
\end{abstract}

\begin{IEEEkeywords}
\rev{GUIOpsBench, cloud operations, GUI agents, vision-language models}
\end{IEEEkeywords}

\section{Introduction}
Modern cloud services are operated through a continuous stream of actions, which typically fall into two broad classes~\cite{zhang2024surveyaiopsfailuremanagement, avrukin2015sreuis}. One is \textit{fault-driven} handling, triggered when incidents occur (e.g., anomaly detection~\cite{logad} and root-cause localization~\cite{nezha})~\cite{zhang2024surveyaiopsfailuremanagement}. The other is \textit{routine} operational work carried out day to day (e.g., resource provisioning, network policy adjustment, and configuration changes)~\cite{avrukin2015sreuis}. Relying entirely on humans to carry out these tasks incurs high labor and time costs, which makes automation desirable. Prior work has largely focused on automating the former. A large body of Artificial Intelligence for Cloud Operations (AIOps) work leverages ML models~\cite{ma2020diagnosing,zhou2018fault} or LLM agents~\cite{chen2023automatic,rcagent,flowofaction} to analyze telemetry (i.e., trace, log, metric) and call APIs for fault diagnosis. In contrast, automating routine operations has received far less attention.

\begin{figure}[t]
    \centering
    \includegraphics[width=1\linewidth]{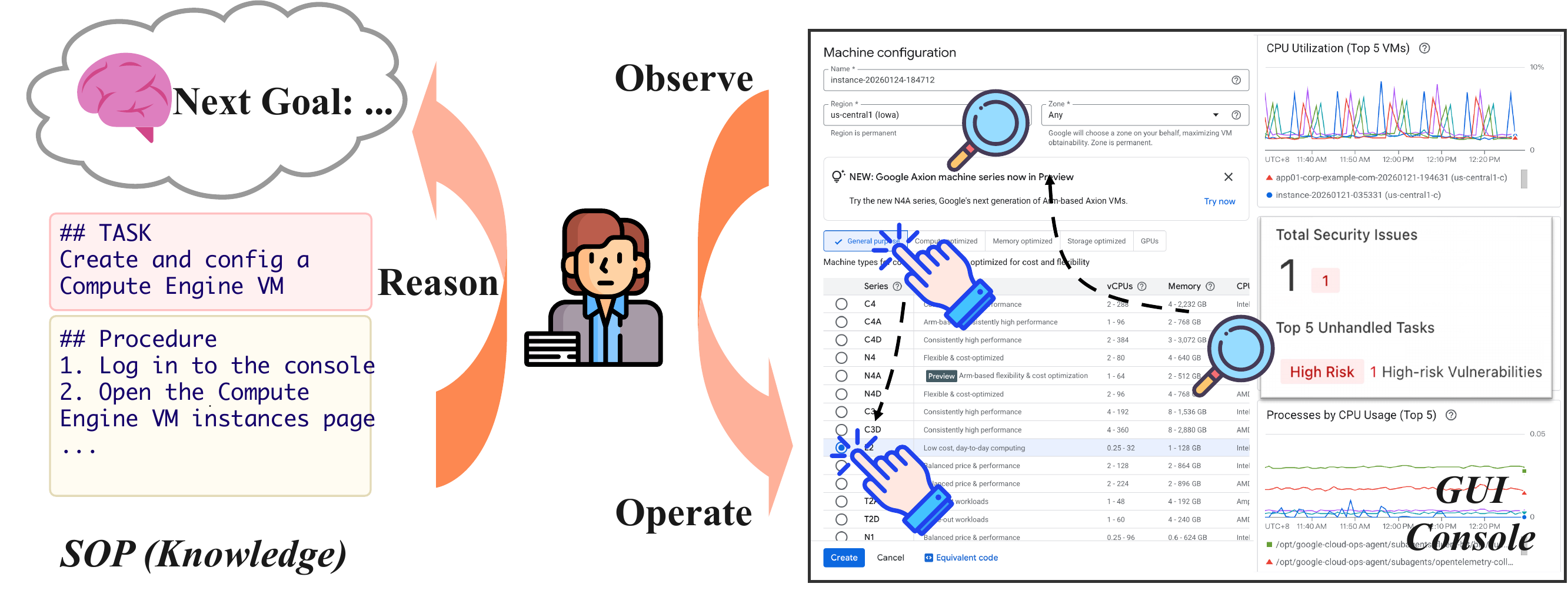}
    \caption{The process of GUI-centric Cloud Operation.}
    \label{fig:background}
\end{figure}

What distinguishes routine Ops tasks from fault-driven tasks most is that they depend on the graphical cloud console. As illustrated in Fig.~\ref{fig:background}, such operations form an observe--reason--operate loop guided by Standard Operating Procedures (SOPs), and are GUI-centric by nature, so prior telemetry-based methods do not apply. Our case study (Sec.~\ref{sec:motivation}) indicates that this is a property of the tasks themselves. First, routine tasks create, delete, or reconfigure resources and thus directly change the production system, so their execution must stay observable and interruptible by a human, for which a console is better suited than a raw API. Second, they often require continuous inspection of dashboards (e.g., Grafana), whose topology graphs, heatmaps, and configuration panels are conveyed more naturally through visuals than through text. Moreover, many console operations still have no programmatic counterpart. In our study, \rev{89\% of the recorded operation tickets were completed entirely through the GUI} (Sec.~\ref{sec:motivation}).

Automating routine GUI tasks is valuable. These tasks are frequent and highly repetitive, and they consume a large share of operators' time. The Google SRE handbook reports that Site Reliability Engineers (SREs) spend roughly 33\% of their time on such toil~\cite{beyer2016sre}, and industry surveys note a strong appetite to automate it~\cite{catchpoint2025srereport}. Consistent with this, in our survey of 20 SREs (Sec.~\ref{sec:motivation}), 90\% rated automating routine GUI tasks as at least as necessary as automating fault-driven ones.

Automating routine GUI tasks calls for agents that can perceive, operate, and verify tasks directly in cloud consoles, which recent advances in vision-language models (VLMs)~\cite{gpt4o,claude-4.5,llama4,bai2025qwen25vltechnicalreport} make increasingly plausible. Yet existing GUI-agent evaluations are largely built around generic web automation that relies on common-sense patterns~\cite{deng2023mind2webgeneralistagentweb,koh2024visualwebarenaevaluatingmultimodalagents,yao2023webshopscalablerealworldweb}, rather than the domain-specific, chart-heavy, SOP-driven nature inherent in cloud operations. Meanwhile, existing cloud-operations benchmarks remain text- and API-centric and do not test GUI interaction~\cite{chen2024aiopslab,jha2025itbenchevaluatingaiagents,DBLP:conf/sigsoft/LiuPXCSZSZW0LXW25}. Therefore, a vision-centric benchmark tailored to cloud management consoles is needed to rigorously evaluate whether agents can execute and verify real Ops procedures through GUIs.

Constructing such a benchmark faces two key challenges: (1) expert operational trajectories are scarce and expensive, as cloud-console tasks require domain expertise that is difficult to crowdsource; and (2) expert demonstrations alone cannot cover the vast configuration space and UI states of consoles, limiting diversity.

To tackle these challenges, we propose a dual-path collection engine. The \textbf{SOP-driven forward pipeline} instantiates Task--SOP pairs from official documentation and collects verified expert-like trajectories. Complementarily, the \textbf{exploration-driven reverse pipeline} performs heuristic-augmented exploration and reconstructs semantic labels post hoc, thereby expanding coverage and increasing data volume.

\begin{figure}[t]
    \centering
    \includegraphics[width=\linewidth]{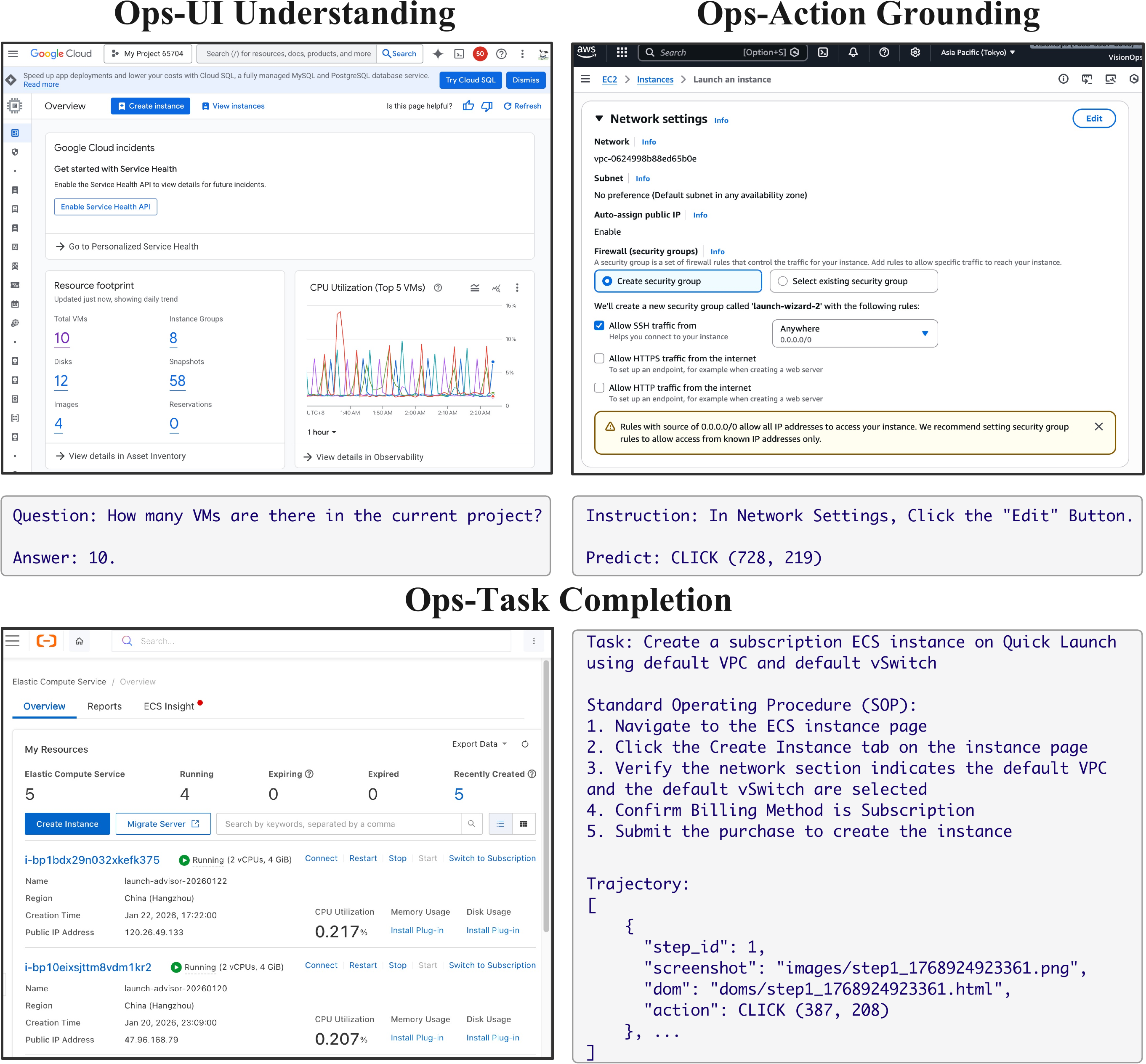}
    \caption{An illustration of three task families in \benchname{}.}
        \label{fig:tasks}
\end{figure}

Leveraging this infrastructure, we construct \textit{\textbf{\benchname{}}}, the first GUI-centric cloud operation benchmark, which contains \textbf{21,680} annotated samples collected from management consoles of four major cloud providers (AWS~\cite{aws}, Google Cloud~\cite{gcloud}, Alibaba Cloud~\cite{alicloud}, and Volcengine~\cite{volcengine}). The dataset has a total storage size of \textbf{101.86 GB} and covers three task families (Fig.~\ref{fig:tasks}): \textbf{Ops-UI Understanding} (15,284 samples), \textbf{Ops-Action Grounding} (5,033 samples), and \textbf{Ops-Task Completion} (1,363 samples). We ensure data quality by introducing an LLM-based judge agent and conducting human reviews by six senior cloud experts.

\rev{We evaluate seven proprietary and open-source VLMs and derive three key findings. First, models struggle with domain-specific operational understanding: Seed-1.6~\cite{doubao16} has an F1 of 0.731 on navigation but 0.196 on specific operations. Second, vision-only task execution remains challenging: GPT-5.2~\cite{gpt52} achieves the highest average task success rate of 13.7\%, and all models' average grounding accuracies are below 20\%. Third, the DOM-assisted TrajMAS framework reaches a highest average task success rate of 50.1\%; Claude-Sonnet-4.5 improves from 2.4\% to 48.9\%, a 20.4$\times$ increase over its vision-only baseline.}

\textbf{Contribution.} Our contributions are three-fold:

\begin{itemize}[leftmargin=*, topsep=0pt, parsep=0pt]
    \item We introduce \textit{\textbf{\benchname{}}}, the first public, GUI-centric benchmark framework tailored to real-world cloud management console operations, covering three task families (Ops-UI Understanding, Ops-Action Grounding, Ops-Task Completion) with 21,680 curated and annotated samples from 4 major cloud providers.
    \item We propose a cross-platform, dual-path data collection engine that scales realistic GUI-centric Ops data via an SOP-driven forward pipeline for high-quality long-horizon trajectories and an exploration-driven reverse pipeline for augmenting data volume and expanding coverage beyond SOP-documented states.
    \item We provide a systematic evaluation of modern VLMs on GUI-centric cloud operations and derive actionable insights on the bottlenecks and effective signals for building reliable autonomous SRE agents.
\end{itemize}

\input{motivation}

\section{Related Work}

\textbf{Benchmarks in Cloud Operations.}
As cloud infrastructure scales, AI-driven cloud operations (AIOps) has emerged as a critical area for automating monitoring, diagnosis, and remediation~\cite{zhang2024surveyaiopsfailuremanagement}.
Troubleshooting benchmarks like TrainTicket~\cite{zhou2018fault} and AIOpsLab~\cite{chen2024aiopslab}, focus on root-cause analysis from logs and metrics.
OpenRCA~\cite{xu2025openrca} provides 335 real failures with over 68\,GB of telemetry for LLM-based root-cause localization, yet is limited to structured telemetry (logs, metrics, traces) rather than console interaction.
OpsEval~\cite{DBLP:conf/sigsoft/LiuPXCSZSZW0LXW25} benchmarks LLMs on IT operations knowledge via multi-choice and QA formats, but is a static knowledge test that does not require agents to act in live environments.
Automation benchmarks such as IaC-Eval~\cite{kon2024iaceval}, ITBench~\cite{jha2025itbenchevaluatingaiagents}, and Ansible-Wisdom~\cite{pujar2023ansiblewisdom} evaluate code generation for infrastructure changes.
On the agent side, LLM-powered approaches such as RCACopilot~\cite{rcacopilot}, Flow-of-Action~\cite{flowofaction}, mABC~\cite{mabc}, and RCAgent~\cite{rcagent} have demonstrated the potential of autonomous agents for cloud incident diagnosis and root-cause analysis; however, these methods operate through APIs and telemetry rather than graphical consoles.
In short, these benchmarks and agents concentrate on fault-driven, telemetry- and API-based tasks, leaving routine GUI-centric operations, the focus of this work, largely untested.

\textbf{GUI Agents and Datasets.}
VLMs have enabled agents to automate tasks across web, mobile, and desktop interfaces~\cite{gpt4o,claude-4.5,llama4,bai2025qwen25vltechnicalreport}.
Recent VLMs such as Qwen3-VL~\cite{bai2025qwen3vltechnicalreport} and InternVL~\cite{chen2024internvlscalingvisionfoundation} add high-resolution input and GUI-oriented training, advancing visual perception and grounding for interactive interfaces.
Web benchmarks such as WebShop~\cite{yao2023webshopscalablerealworldweb}, Mind2Web~\cite{deng2023mind2webgeneralistagentweb}, and VisualWebArena~\cite{koh2024visualwebarenaevaluatingmultimodalagents} focus on navigation and e-commerce;
mobile benchmarks like AITW~\cite{rawles2023androidwildlargescaledataset}, RICO~\cite{10.1145/3126594.3126651}, and AitZ~\cite{zhang2024androidzoochainofactionthoughtgui} cover Android apps;
desktop benchmarks such as OfficeBench~\cite{wang2024officebenchbenchmarkinglanguageagents} and GUI-360~\cite{mu2025gui360circcomprehensivedatasetbenchmark} target office software.
Cross-platform OSWorld~\cite{xie2024osworld} evaluates agents on 369 open-ended tasks across real OSes, while enterprise-oriented WorkArena~\cite{drouin2024workarena} targets knowledge-work tasks in ServiceNow.
While diverse, these benchmarks address general-purpose tasks and lack the domain-specific semantics, SOP-driven procedures, and safety constraints that characterize cloud operations.

\section{Problem Formulation}

We formalize GUI-centric cloud operations as a partially observable decision process with procedural constraints.

\textbf{Operations Environment.}
We define the state space $\mathcal{S}$ with $s_t = (I_t, D_t)$ comprising a console screenshot $I_t$ and DOM tree $D_t$, the GUI action space $\mathcal{A}$ with $a = (\textit{type}, \textit{target}, \textit{value})$ for $\textit{type} \in \{\texttt{click}, \texttt{type}, \texttt{scroll}$, $\texttt{wait}\}$, and $\mathcal{G}$ as the set of operational goals.

\textbf{Standard Operating Procedures (SOPs).}
Cloud operations are procedure-driven. For each goal $g \in \mathcal{G}$, a Standard Operating Procedure $\textit{SOP}_g = \langle \mathcal{P}, \mathcal{K}, \mathcal{V} \rangle$ specifies prerequisite conditions $\mathcal{P}$, an ordered sequence of atomic steps $\mathcal{K} = (k_1, \ldots, k_n)$, and a verification function $\mathcal{V}: \mathcal{S} \rightarrow \{0, 1\}$ that determines task completion from the terminal.

\textbf{Trajectory and Task Completion.}
Given goal $g$ and initial state $s_0$, an operational trajectory is a sequence $(s_0, a_0, s_1, \ldots$, $a_{T-1}, s_T)$. Task success is determined by $\mathcal{V}_g(s_T) \in \{0, 1\}$.

\textbf{Operation Cycle.}
An agent executes tasks through an iterative \emph{Perceive--Reason--Act} cycle. At each step $t$, it \textbf{perceives} $s_t$ and aligns it with the current SOP step $k_i$; \textbf{reasons} about the state transition needed to satisfy $k_{i+1}$; and \textbf{acts} by grounding the instruction into a concrete GUI action $a_t$, yielding $s_{t+1}$.

\begin{figure*}[t]
    \centering
    \preprintgraphic[width=\linewidth]{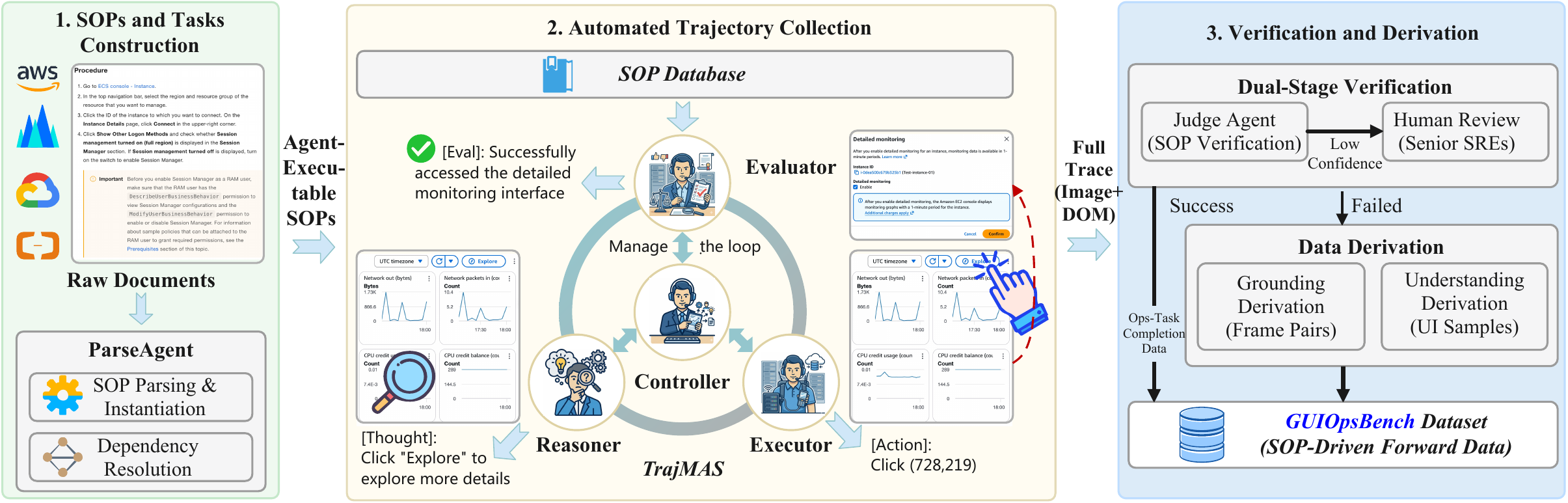}
    \caption{The overall architecture of SOP-Driven Forward Pipeline.}
    \label{fig:forward}
    \end{figure*}

\section{\rev{Dataset Construction}}
Building GUI-centric operations datasets faces two key challenges. \textbf{\textcircled{\raisebox{-0.9pt}{1}} Scarcity and high cost of expert data:} Real-world operations require rich context and domain expertise, which cannot be easily crowdsourced like general web tasks. \textbf{\textcircled{\raisebox{-0.9pt}{2}} Coverage and limited exploration space:} Forward-constructed data is inherently bounded by the operational space documented in SOPs, leaving vast UI states and configurations unexplored.

To address these challenges, we propose a novel cross-platform, dual-path data collection engine with two complementary pipelines. \textbf{The SOP-Driven Forward Pipeline} uses an SOP-guided agent to generate high-quality, long-horizon expert trajectories. \textbf{The Exploration-Driven Reverse Pipeline} employs an explorer agent to perform heuristic-enhanced automated random walks and semantic reconstruction, expanding data coverage and scale.

\subsection{The SOP-Driven Forward Pipeline}
The forward pipeline uses agents to simulate experienced SREs and automatically produce ``gold-standard'' data meeting industrial standards. It follows a three-stage design to maximize scalability while minimizing human effort, as shown in Fig.~\ref{fig:forward}. (1) \textbf{SOPs and tasks construction.} We design an \textit{ParseAgent} to crawl, clean, and instantiate real-world operation tasks and related SOPs from official documentation of four major cloud providers. (2) \textbf{Automated trajectory collection.} We build a specialized multi-agent system, called \textit{TrajMAS}, to execute tasks automatically while recording the full execution trajectory. (3) \textbf{Verification and derivation.} We introduce a dual verification process to validate collected trajectories, from which final data points in \benchname{} are then derived.

\subsubsection{SOPs and Tasks Construction}
\begin{figure}[t]
    \centering
    \includegraphics[width=\linewidth]{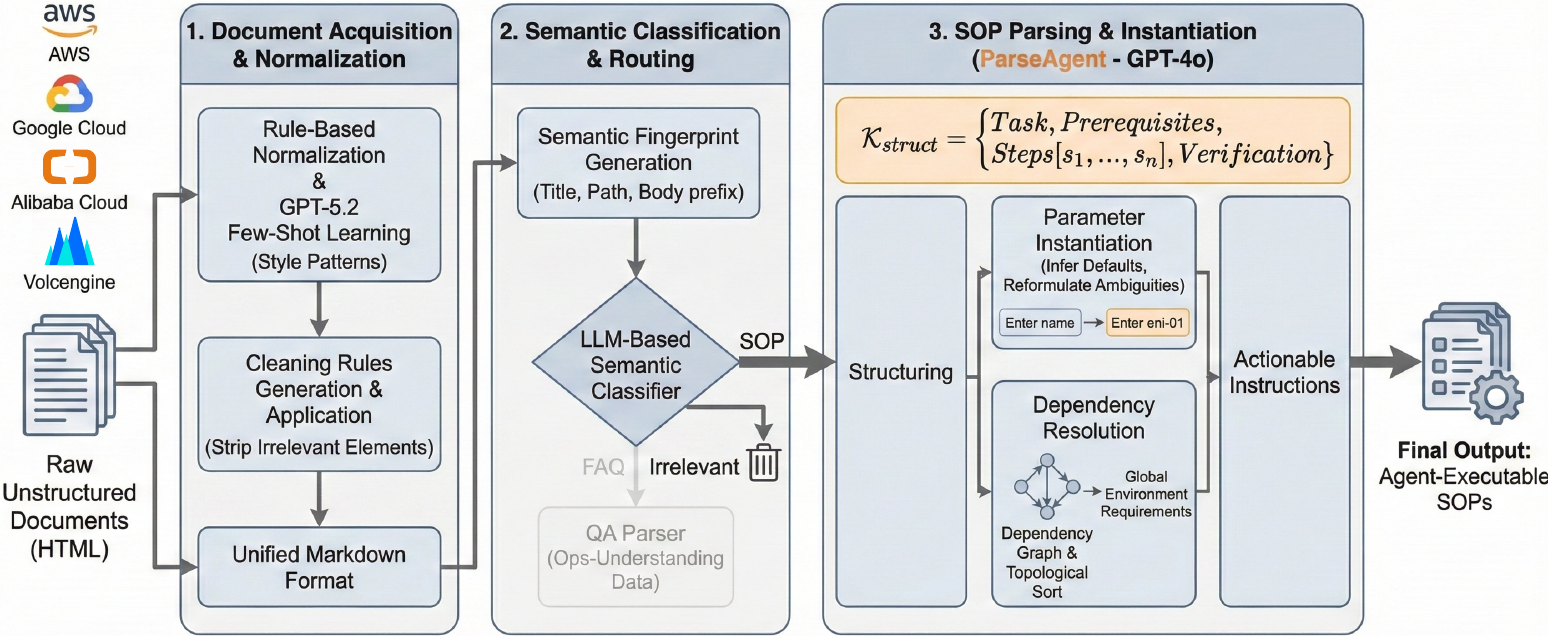}
    \caption{The process of SOPs and tasks construction.}
        \label{fig:sop_construct}
\end{figure}

Cloud operations are inherently process-dependent; a slight deviation from the standard procedure can lead to critical failures. Therefore, high-quality, actionable SOPs are the cornerstone of our benchmark. We collected and structured 624 real-world Task--SOP pairs from the latest official documentation of four major cloud providers (AWS~\cite{aws}, Google Cloud~\cite{gcloud}, Alibaba Cloud~\cite{alicloud}, and Volcengine~\cite{volcengine}). The process of SOPs construction is shown in Fig.~\ref{fig:sop_construct}.

\textbf{Document acquisition and normalization.} Raw documentation from various cloud providers is unstructured (HTML) and inconsistent in style. To standardize these inputs, we first applied a rule-based normalization process. For each platform, we utilized GPT-5.2 to learn platform-specific styling patterns from few-shot examples (3-5 documents) and generate cleaning rules to strip irrelevant elements (e.g., sidebars, promotional banners) while preserving the procedural core. The cleaned content was then reorganized into a unified Markdown format. 

\textbf{Semantic classification and routing.} Because documents serve different purposes, they are processed differently. For example, ``best practice'' documents are suitable for extracting SOPs, while ``overview'' documents are better for QA. We therefore route documents to different parsers. To route documents effectively, we generated a Semantic Fingerprint for each file, comprising its title, document path, and the  500 characters summary of the body. An LLM-based semantic classifier then categorized each document into three types: SOP, FAQ, or Irrelevant. Documents labeled as SOP were routed to the SOP Parser, while FAQ documents were processed by a QA Parser to extract auxiliary QAs for Ops-Understanding task.

\textbf{SOP parsing and instantiation.} We designed a ParseAgent (powered by GPT-4o) to transform raw SOP documents into a structured format containing four fields: \textit{Task} (the high-level operation goal), \textit{Prerequisites} (required preconditions), \textit{Steps} (a sequence of atomic actions with expected intermediate states), and \textit{Verification} (terminal state criteria). Crucially, raw documentation is often underspecified (e.g., ``Configure relevant parameters"). To make these SOPs agent-executable, ParseAgent performs \textbf{Parameter Instantiation}: it infers valid default values based on field descriptions (e.g., converting ``Enter a name" to ``Enter eni-01") and reformulates ambiguous steps into actionable instructions. Furthermore, to resolve \textbf{Prerequisite Dependencies} (e.g., creating a volume requires an existing snapshot, which may be related to another SOP on snapshot creation), we constructed a dependency graph of all SOPs. We performed a topological sort to execute upstream dependencies first. For unresolved external dependencies, we aggregated their prerequisites into a global environment requirement set, ensuring the sandbox environment was pre-provisioned before execution.

\subsubsection{Automated Trajectory Collection}
\label{subsec:TrajMAS}
With instantiated SOPs ready, the next challenge is execution. Manual collection is prohibitively expensive and unscalable. To address this, we developed a specialized multi-agent system, \textit{TrajMAS} (Trajectory Multi-Agent System), to automate data collection in a quality-ensured manner.

\textbf{Execution Environment.} To bridge the gap between simulation and reality, we conducted all data collection on live production consoles of the target cloud providers. We utilized isolated namespaces and resource groups to prevent interference. High-risk actions (e.g., account deletion) were gated by secondary authentication, allowing agents to navigate the UI flow without triggering destructive consequences. The session is initiated via a headless browser (Playwright), where human operators handle the initial login, after which control is fully handed over to \textit{TrajMAS}.

\textbf{\textit{TrajMAS} Architecture.} \textit{TrajMAS} is a multi-agent system designed for addressing GUI-centric cloud operation tasks, which operates as a closed-loop system comprising four agents (all powered by GPT-5.2), as shown in Fig.~\ref{fig:forward}.

The \textbf{Controller} fetches a task from the SOP library, manages the Perception-Reasoning-Action-Evaluation loop, maintains the global state, and decides when to terminate the session. The \textbf{Reasoner} aligns the current screenshot with the active SOP step and infers the next GUI operation (Click, Type, Scroll, Wait); to improve precision and reduce trial-and-error, it applies Set-of-Mark (SoM) prompting, leveraging the Chrome DevTools Protocol (CDP) to extract bounding boxes of interactive elements so that it outputs element indices (e.g., Click(`Create[idx:1]')) rather than error-prone absolute coordinates. The \textbf{Executor} runs the inferred actions via Playwright and records multimodal data, including high-resolution screenshots, DOM trees, and agent outputs. The \textbf{Evaluator} compares the pre- and post-action states to generate a summary, assesses whether the current SOP step is completed, and updates the task progress.

The loop continues until the Evaluator marks the task as complete or a maximum of 20 steps is reached. TrajMAS then ends current SOP, saves full trajectory in JSON format, resets to root console URL, and proceeds to next SOP.

\subsubsection{Data Verification and Derivation}
\textbf{Dual-Stage Verification.} To ensure trajectory quality and consistency with SOPs, we apply dual-stage verification. After each trajectory is collected, a Judge Agent evaluates the SOP verification function $\mathcal{V}_g(s_T)$ by comparing the final observed state against the expected outcome, and emits a ternary verdict: \textit{Definite Pass}, \textit{Ambiguous}, or \textit{Definite Fail}. We adopt a conservative trust policy: only trajectories receiving a \textit{Definite Pass} verdict are automatically accepted. All remaining cases (both \textit{Ambiguous} and \textit{Definite Fail}) are escalated to 6 senior SREs for manual review. This design reflects our observation that LLM-based judges are prone to specific failure modes, e.g., misclassifying a payment confirmation page as task failure, or prematurely accepting a task when domain-specific keywords appear on a landing page despite incomplete execution. By trusting the Judge Agent only when it is highly confident, we maximize automation while safeguarding label accuracy. Statistically, 53\% of collected trajectories passed this rigorous verification and were labeled as Success.

\textbf{Data Derivation.} For data efficiency, successful trajectories become high-quality samples for the Ops-Task Completion dataset, while failed trajectories are not discarded but labeled as weak-semantic trajectories and reused as raw data to derive Understanding and Grounding tasks. A single agent run can thus produce data for all downstream tasks through derivation. \textcircled{\raisebox{-0.9pt}{1}} Understanding Derivation. The state summaries generated by the Evaluator Agent are paired with their corresponding screenshots to derive Ops-UI Understanding samples 
(e.g., ``Describe the content of the heatmap in the red box").
\textcircled{\raisebox{-0.9pt}{2}} Grounding Derivation. Every adjacent frame pair $(s_t, a_t, s_{t+1})$ in a trajectory is decomposed into an atomic Grounding sample: the initial screenshot $S_t$ serves as the input, the parsed instruction serves as the query, and the executed element's coordinates in $a_t$ serve as the ground truth label. The quality of derived samples is inherently guaranteed: the ground-truth coordinates and screenshots come directly from actions actually executed on live consoles, rather than from model predictions or human annotations. This turns a modest set of SOP documents into a large, multi-faceted dataset.

\subsection{The Exploration-Driven Reverse Pipeline}
While the Forward Pipeline ensures high-quality, task-oriented data, it is inherently constrained by the coverage of existing documentation, and cloud consoles contain many elements and configurations that are rarely documented in standard SOPs. To bridge this coverage gap, we introduce the Exploration-Driven Reverse Pipeline. Unlike the forward approach, it operates without predefined tasks, adopting an ``Explore-First, Attribute-Later'' paradigm whose explicit goal is expanding state-space coverage and data volume rather than completing any specific operational task. It autonomously traverses the console to collect digital breadcrumbs and subsequently reconstructs semantic labels via inverse dynamics.

\begin{figure}[t]
    \centering
    \preprintgraphic[width=\linewidth]{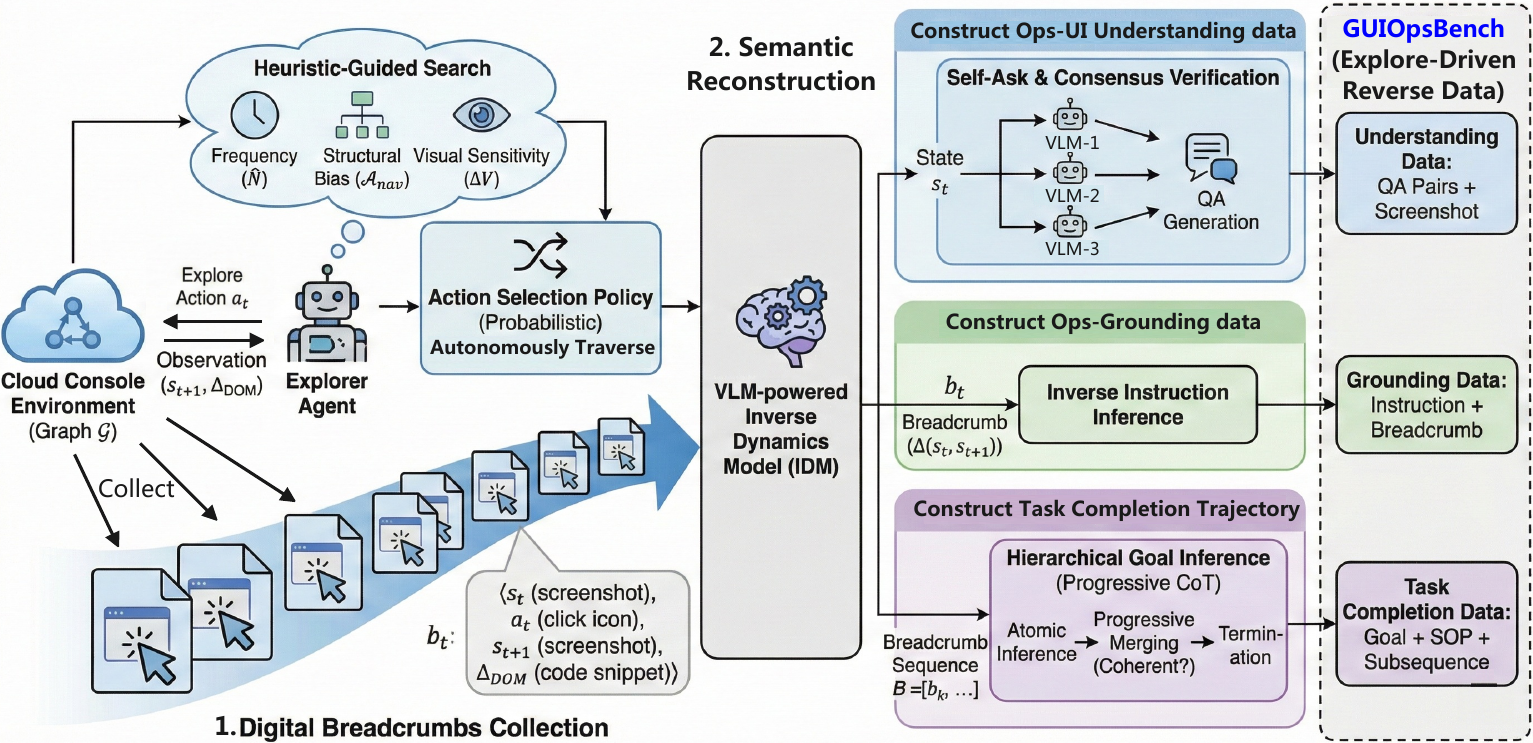}
    \caption{The process of Exploration-Driven Reverse Pipeline.}
        \label{fig:reverse}
\end{figure}

\subsubsection{Heuristic-Guided GUI Exploration}
Navigating a cloud console can be modeled as traversing a massive directed graph $G = (V, E)$, where nodes $V$ are unique GUI states (i.e., pages, snapshots) and edges $E$ are actionable transitions. 
Fully random walking in such a high-dimensional space is notoriously inefficient, often trapping agents in loops or empty states. To address this, we develop an Explorer Agent based on a heuristic-augmented search algorithm that preserves randomness while guiding exploration more intelligently, preventing repeated actions and frequent dead ends. Crucially, the Explorer Agent's sole objective is to expand state-space coverage for data generation; it does not attempt to complete any operational task or follow any predefined goal. Because the exploration is driven entirely by visit-frequency and structural heuristics rather than by a model's task-completion policy, the resulting data is free from the inductive bias of any particular VLM or agent strategy.

\textbf{Action Potential Estimation.} A critical challenge in GUI exploration is that the destination URL of an action $a$ is often unknown until it is executed (e.g., JavaScript-driven transitions). To handle this, the agent maintains a state-action memory $M(s, a) \rightarrow s'$ and a global state visitation counter $N(s)$ for every unique state $s$ visited. We define the Estimated Destination Frequency $\hat{N}(s_t, a)$ for an action $a$ in current state $s_t$ as:
\begin{equation}
\hat{N}(s_t, a) = \begin{cases} 0 & \text{if } a \text{ is never triggered } \\ N(M(s_t, a)) & \text{if } a \text{ has been triggered } \end{cases}
\end{equation}
This formulation ensures that unexplored actions are treated as leading to new states, thereby receiving the highest exploration priority. 

\textbf{Heuristic Search Strategy.} 
The Explorer Agent selects the next action $a_t \in \mathcal{A}$ using a set of heuristic rules that aim to maximize state coverage while preferentially exploring informative pages. (1) Low-Frequency Bias favors actions that lead to rarely visited states, with the score decaying as the estimated visit count $\hat{N}$ increases. (2) Structural Bias exploits DOM analysis to identify the console's sidebar navigation tree, upweighting actions $a \in \mathcal{A}_{nav}$ to ensure coverage of diverse service modules. (3) Visual Sensitivity applies a visual difference filter, where an action is considered valid only if the visual change $\Delta V(s_t, s_{t+1}) > 0$, measured by perceptual hash distance; actions that fail to trigger visual changes are recorded as invalid ($\mathcal{A}_{inv}$) and excluded from future exploration. The priority score for each action is then computed as:
\begin{equation}
    H(a, s_t) = \frac{\lambda_{freq}}{\sqrt{\hat{N}(s_t, a) + 1}} + \lambda_{nav} \cdot \mathbb{I}(a \in \mathcal{A}_{nav}),
\end{equation}
where  $a \notin \mathcal{A}_{inv}$ and $\lambda$  denotes hyperparameters that balance the heuristics.

\textbf{Transition Probability.} The transition probability $P(a|s_t)$ in the Explorer Agent is inspired by the damping factor in the PageRank algorithm, which allows random jumps to prevent getting trapped in sink nodes. We model the policy as a mixture of a heuristic-guided distribution and a uniform distribution:
\begin{equation}
P(a|s_t) = d \cdot \frac{e^{H(a, s_t)/\tau}}{\sum_{a' \in \mathcal{A}} e^{H(a', s_t)/\tau}} + (1 - d) \cdot \frac{1}{|\mathcal{A}|}
\end{equation}
where $d \in [0, 1]$ is the damping factor controlling the balance between following the heuristic score and random wandering, and $\tau$ is a temperature parameter controlling the sharpness of the softmax distribution. In our experiments, we set $d = 0.85$ and $\tau = 1.0$. This probabilistic formulation ensures that while the agent strongly prefers unvisited or high-value actions, it always retains a non-zero probability of choosing arbitrary actions, preventing cycles and ensuring asymptotic coverage of the state space.

\subsubsection{Collecting Digital Breadcrumbs}
During exploration, the Explorer Agent leaves behind a trail of what we term ``Digital Breadcrumbs'', i.e., transition tuples $b_t = \langle s_t, a_t, s_{t+1} \rangle$ that record pre-/post-action screenshots and doms, and the executed action. Although each breadcrumb lacks explicit user intent, it encodes ground-truth environmental dynamics and forms the raw substrate for downstream semantic reconstruction.

\subsubsection{Semantic Reconstruction}
To transform these raw breadcrumbs into labeled data required for \benchname{}, we employ a VLM-powered agent as an Inverse Dynamics Model (IDM). The IDM performs post-hoc semantic attribution, inferring the latent intent that would have motivated the observed transitions. Unlike the forward pipeline, where intent precedes execution and requires post-hoc verification, the reverse pipeline records real executed transitions as ground truth. The action coordinates, DOM mutations, and state changes in each breadcrumb are deterministic and factual, so the IDM-generated semantic labels (instructions, goals) can be programmatically validated against these recorded transitions.

\textbf{Construct Ops-UI Understanding Data.} We leverage the self-ask capability of agent to generate Understanding samples. For each unique state $s_t$ in the breadcrumbs, the IDM analyzes $s_t$ (image and DOM) and generates $K=8$ distinct QA pairs about operational functions (e.g., ``Which button should I click to recharge my account?''). Each answer includes a natural-language sentence and a list of answer keywords. To reduce hallucination and prevent single-model bias, the agent answers each question three times using diverse backbones from different vendors (i.e., GPT-5, Gemini-3-Flash, Qwen3), serving as a cross-validation mechanism.
A QA pair is accepted only if keyword lists match across all three answers. 
Each accepted QA, together with its associated screenshot, forms one datapoint in the Understanding dataset.
To further validate the quality of all LLM-derived Understanding labels (from both the forward and reverse pipelines), we randomly sampled 20\% of the generated QA pairs for manual expert review.
The review confirmed an agreement rate of \textbf{96\%} between the LLM-generated labels and human expert judgments, demonstrating that the annotation error rate is well controlled.

\textbf{Construct Ops-Grounding Data.} We reconstruct grounding instructions by solving the inverse problem: `given the state change, what was the instruction?' For each breadcrumb $b_t$, the IDM is prompted with the visual difference $(s_t, s_{t+1})$ and asked to reconstruct the likely high-level instruction. For example, if $s_t$ shows an instance list and $s_{t+1}$ displays a dialog saying ``Confirm restart eni-1,'' the VLM infers the instruction ``Click the restart button of instance eni-1.'' The executed action $a_t$ and its coordinate become the ground-truth action, and $s_{t+1}$ is the expected post-action state. Each inferred instruction, together with its breadcrumb $b_t = \langle s_t, a_t, s_{t+1} \rangle$, forms one datapoint in the Grounding dataset.

\textbf{Construct Task Completion Trajectories.} Synthesizing multi-step tasks from random walks is the most challenging component. We employ a hierarchical goal inference strategy. Let $B = [b_k, b_{k+1}, ..., b_{k+n}]$ be a continuous sequence of breadcrumbs. The IDM analyzes this sequence using a progressive Chain-of-Thought (CoT) approach to synthesize a high-level goal: (1) Atomic Inference: First, infer the intent of $b_k$. (2) Progressive Merging: Iteratively attempt to merge $b_{k+1}$ into the narrative. The model asks: "Can actions $a_k$ and $a_{k+1}$ be explained by a single coherent goal?" (3) Termination: The sequence grows until the IDM determines that adding $b_{k+m}$ disrupts the logical coherence of the task. The valid subsequence $B_{1..m}$ is then labeled with the synthesized High-Level Goal (e.g., ``Navigate to the Virtual Private Cloud (VPC) dashboard and check security group settings"). Finally, the IDM generates a corresponding SOP based on the observed actions. Thus, a datapoint in Task Completion dataset is constructed from a semantic-free breadcrumb sequence.

\textbf{Coverage Expansion.}
The heuristic-enhanced exploration discovered \textbf{7,330} unique GUI states that were never visited by the forward pipeline, reaching console areas that standard SOPs do not cover---for example, billing management dashboards, scheduled-task configuration panels, and advanced quota/limit pages.
This expansion is by design: the Explorer Agent's frequency-based and structural heuristics steer it toward rarely visited navigation branches and deep sub-menus, complementing the SOP-driven data with broader state-space coverage.

\textbf{Human Verification of Reverse Trajectories.}
For the Task Completion trajectories synthesized from breadcrumb sequences, we conducted a rigorous post-hoc human review.
Six senior SREs independently examined each synthesized trajectory for semantic correctness, logical coherence, and operational meaningfulness.
Trajectories that were semantically incoherent, logically erroneous, or operationally trivial (e.g., repetitive back-and-forth navigation without achieving a meaningful state change) were discarded.
After this review, \textbf{68\%} of the synthesized trajectories were retained, resulting in 739 reverse-constructed Task Completion samples.

\textbf{On the Reliability of Reverse-Pipeline Labels.}
A potential concern is whether reverse-pipeline annotations amount to free-form model imagination.
Here, each label is \emph{post-hoc} constrained by verifiable breadcrumbs---the executed action $a_t$, pre/post observations $(s_t,s_{t+1})$, and the resulting DOM mutation $\Delta\mathrm{DOM}$---and by \emph{visibility constraints} that only allow formulations supported by what is observable in $s_t$.
Accordingly, we treat reverse-pipeline outputs as a \emph{plausible reconstruction} of operational intent, and evaluate \emph{operationally actionable understanding} under visible UI evidence and executable action constraints rather than latent user psychology.

\begin{figure}[t]
    \centering
    \preprintgraphic[width=\linewidth]{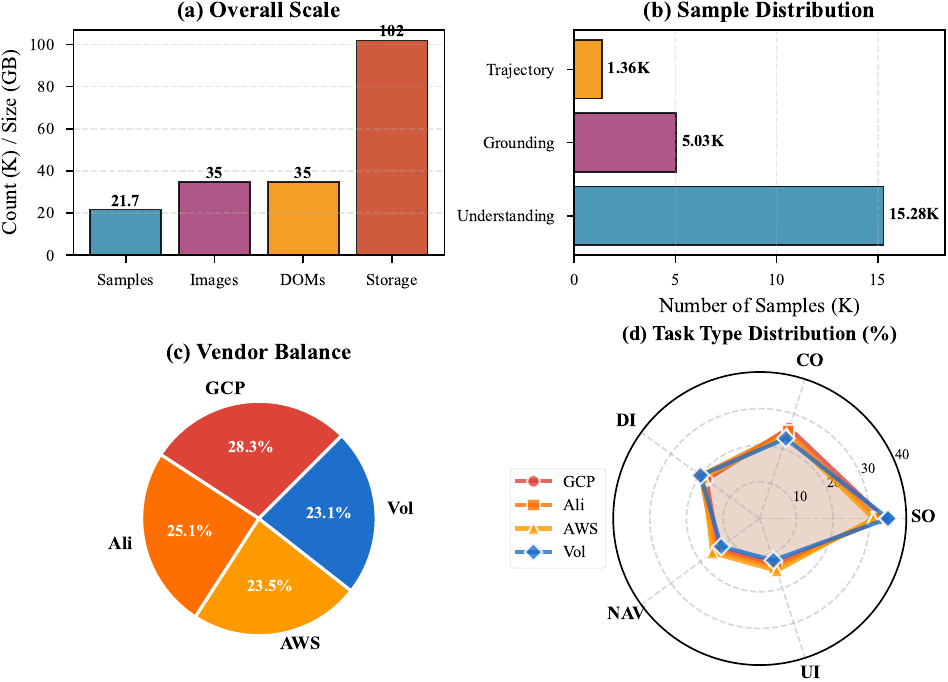}
    \caption{The statistic of \benchname{} (GCP denotes Google Cloud, Ali denotes Alibaba Cloud, Vol denotes Volcengine, SO denotes Specific Operations, CO denotes Console Overview, DI denotes Data Interpretation, NAV denotes Navigation, and UI denotes UI element).}
        \label{fig:statistics}
\end{figure}

\subsection{\benchname{} Statistics}

\textbf{Scale.} \benchname{} contains 21,680 samples spanning 13,589 steps, with a total storage size of 101.86\,GB and over 48K screenshots and DOM snapshots. The dataset comprises three complementary data types:
(1) \textit{Understanding} (15,284 samples): Question-answering pairs covering 5 distinct task types (specific operations, console overview, data interpretation, navigation, and UI element understanding), enabling comprehensive multimodal comprehension evaluation.
(2) \textit{Grounding} (5,033 samples): UI element localization annotations with 10K+ before-after screenshot pairs, providing pixel-level supervision for grounding natural language instructions to UI regions.
(3) \textit{Trajectory} (1,363 samples): Complete task execution traces with an average of 9.97 steps, including SOP-driven, forward-constructed demonstrations (624 cases, 45.78\%) and exploration-driven, backward-constructed cases (739 cases, 54.22\%), which together ensure both expert quality and broad coverage.

\textbf{Diversity.} The dataset spans 4 major cloud platforms (AWS, Alibaba Cloud, Google Cloud, Volcengine) with balanced distribution (variance $<$ 6\%). \rev{The Understanding subset covers five categories: Specific Operation, Console Overview, Data Interpretation, Navigation, and UI Element.} The overall statistics are in Fig.~\ref{fig:statistics}.

\section{Evaluation}

We evaluate seven VLMs on \benchname{} across Ops-UI Understanding, Ops-Action Grounding, and Ops-Task Completion. They include five proprietary models: GPT-4o, GPT-5.2, Claude-Sonnet-4.5, Gemini-3-Flash, Seed-1.6, and two open-source, vision-specialized models: Qwen3-VL (235B) and InternVL (7B). For proprietary models, we select the latest mid-tier models from each vendor rather than top-tier models (e.g., Gemini-3-Pro and Claude-Opus), because preliminary tests indicate that top-tier models incur prohibitively high latency and monetary costs (over 35 seconds per step), making them impractical for cloud operation scenarios that require high-frequency interactions.

\subsection{Ops-UI Understanding}

\textbf{Task Definition.} The Ops-UI Understanding task assesses a model's ability to interpret high-density operational GUIs. Given a screenshot $I$ and a domain-specific query $Q$, the model is required to generate a natural language response $R$ and a list of keywords $K_{pred}$ (e.g., component names, status indicators). The task encompasses five sub-categories: console overview, UI element, navigation, specific operation, and data interpretation.

\textbf{Evaluation Metrics.} We evaluate two complementary aspects: (1) \textit{Keyword Matching}: we compute standard Precision, Recall, and F1-score between the predicted keyword list $K_{pred}$ and the ground-truth list $K_{gt}$; (2) \textit{Semantic Similarity}: we measure the cosine similarity between response embeddings using Qwen3-Embedding-8B as encoder.

\begin{figure}[t]
    \centering
    \includegraphics[width=\linewidth]{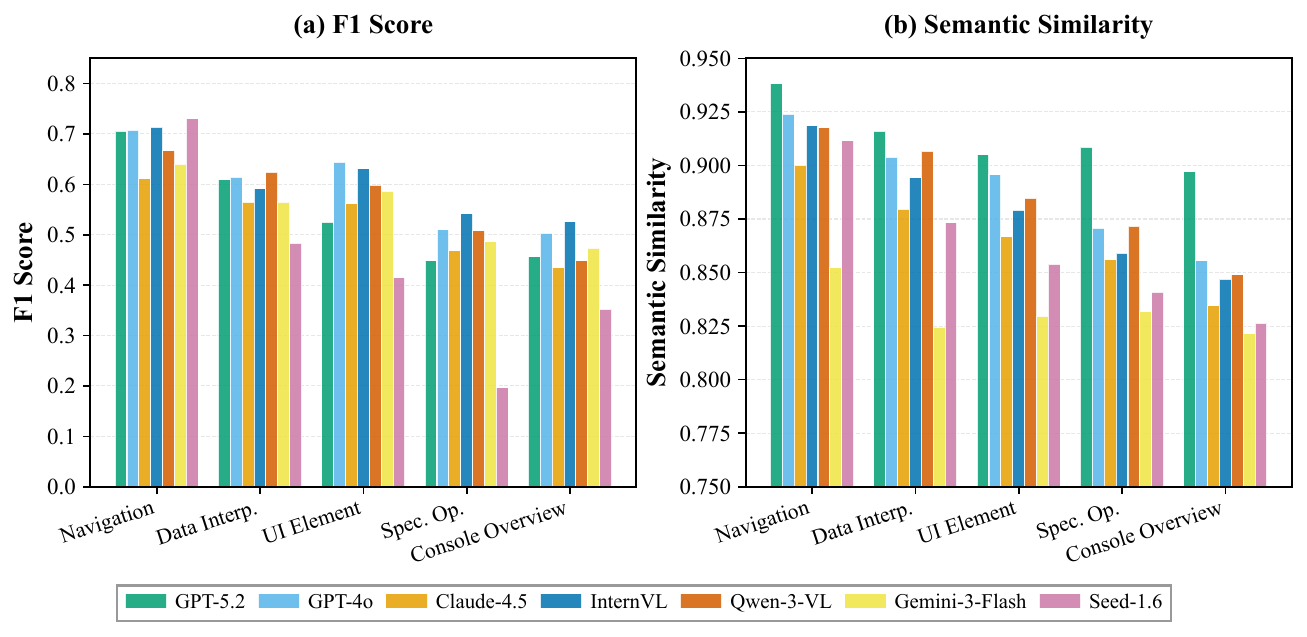}
    \caption{Performance comparison across task types on Ops-UI Understanding.}
        \label{fig:understanding}
\end{figure}

\textbf{Models perform worse on Ops domain-specific understanding tasks.} As shown in Fig.~\ref{fig:understanding}, performance varies across task types. While VLMs are relatively robust on general GUI tasks such as navigation (average F1 of 0.682), their performance degrades on domain-specific tasks. In particular, \textit{Console Overview} (average F1 of 0.456) and \textit{Specific Operation} (average F1 of 0.451) are the most challenging. This indicates that although current VLMs can recognize generic UI elements, they struggle to understand the complex layout logic of Ops consoles and lack the domain knowledge needed to perform specific operations. This limitation is especially evident in some models. For example, although Seed-1.6
achieves top-tier performance on navigation tasks (highest F1 of 0.731), it fails catastrophically on \textit{Specific Operation}, where its F1 drops to 0.196.

\textbf{Comparison between different models.} In terms of semantic similarity, the proprietary model GPT-5.2 outperforms all others across tasks, achieving an average score of 0.913, followed closely by GPT-4o (0.890). However, on keyword F1, GPT-5.2 ranks only fifth, with an average F1 of 0.5487, trailing the open-source models InternVL (0.6006, highest) and Qwen-3-VL (0.569). A closer inspection shows that GPT-5.2 tends to generate many keywords, introducing redundant information that lowers precision and its overall F1 score. This reveals that flagship proprietary models produce more semantically accurate responses, while open-source models generate more concise keyword-level outputs.

\input{tab/grounding}

\input{tab/task-completion}

\subsection{Ops-Action Grounding}
\textbf{Task Definition.}The Ops-Action Grounding task evaluates the agent's fine-grained visual localization capabilities within complex operational interfaces. Given a console screenshot $I$ and a specific action instruction $T$, the model must predict the distinct pixel coordinates $(x, y)$ required to interact with the target element. 

\textbf{Evaluation Metric.} We adopt \textit{Box-Hit Accuracy}: a prediction is correct if and only if the predicted coordinate falls within the ground-truth bounding box of the target element. This is stricter than Euclidean-distance metrics commonly used in prior work, reflecting the narrow interactive margins of real console elements. Formally, let $B_{gt} = [x_{min}, y_{min}, x_{max}, y_{max}]$ denote the ground-truth bounding box and $P_{pred} = (x, y)$ the predicted coordinate; accuracy is $\frac{1}{|\mathcal{D}|} \sum_{i} \mathbb{I}( x_{min}^{(i)} \le x^{(i)} \le x_{max}^{(i)} \land y_{min}^{(i)} \le y^{(i)} \le y_{max}^{(i)})$.

\rev{\textbf{The evaluated VLMs perform poorly on Ops-Grounding tasks.}} As shown in \rev{Table~\ref{tab:grounding_results}}, Ops-Action Grounding is the primary bottleneck in the vision-only autonomous operations pipeline for current VLMs. Overall performance is uniformly low, with no model achieving an average accuracy above 20\%.
During our experiments, GPT-5.2 exhibits noticeably better grounding accuracy and stability than other models, with an average accuracy of 19.09\%, and shows consistent performance across all four vendors. This suggests that it does not overfit to specific page styles and may instead capture a more general grounding strategy. In contrast, all other proprietary and open-source models perform substantially worse, with average accuracies below 8\%. The low accuracy across all backbones indicates that while VLMs may understand what to click at the semantic level, they lack the pixel-level spatial precision needed to map that intent to exact coordinates in software interfaces.

\textbf{Error Analysis.} We further find a prevalent ``Density Dilemma'' failure mode: cloud consoles feature extreme information density with highly repetitive UI components, causing predicted coordinates to drift by a few pixels and miss narrow target bounding boxes. Our qualitative analysis reveals that this near-miss phenomenon is especially common when instructing models to click small targets (e.g., a 16$\times$16 pixel icon) within crowded data tables containing many identical UI elements (e.g., ten ``Edit'' icons in a single column). The predicted coordinate often drifts into adjacent whitespace or a neighboring cell, indicating that current visual encoders suffer from significant information loss when compressing high-resolution inputs (e.g., 4K screenshots) into a limited number of tokens. This finding suggests that relying solely on visual information for Ops tasks is currently insufficient, and structured information (e.g., DOM) should be leveraged to bypass this grounding bottleneck.

\subsection{Ops-Task Completion}
\textbf{Task Setup and Environment.} The Ops-Task Completion task serves as the ultimate litmus test for autonomous SRE agents. Unlike static benchmarks, we evaluate agents in a live, interactive environment powered by Playwright. Agents are granted authorized access to the management consoles of four major cloud providers (AWS, Alibaba Cloud, Google Cloud, Volcengine). Given a high-level goal $g \in \mathcal{G}$ and its standard operating procedure $\textit{SOP}_g$, the agent must navigate the console, interact with UI elements, and execute a sequence of actions to alter the cloud resource state to a target configuration. We employ two distinct frameworks for this evaluation: (1) Vision-Only Baseline (Out-of-the-Box): The agent relies solely on raw screenshots for perception and directly predicts pixel coordinates for action execution, simulating a pure human-like visual interaction. (2) TrajMAS Framework (as described in Sec.~\ref{subsec:TrajMAS}) : The agent operates within our proposed multi-agent system, leveraging the DOM tree and Set-of-Mark (SoM)~\cite{yang2023setofmarkpromptingunleashesextraordinary} prompting to bridge the visual grounding gap.

\textbf{Evaluation Metrics.} We report two execution-based metrics: (1) \textit{Success Rate (SR)}: $SR = \frac{1}{|\mathcal{D}|} \sum_{i} \mathbb{I}\!\left(\text{Verify}\!\left(\text{State}_{final}^{(i)}\right) = \text{True}\right)$, i.e., the fraction of tasks whose final cloud-resource state passes the ground-truth verification; (2) \textit{Trajectory Quality (Pr, Rc, F1)}: step-level alignment between the agent's action sequence and the ground-truth trajectory, where a step matches only if both action type and target element agree.

\textbf{Overall Analysis.} \rev{Table~\ref{tab:task_completion} reports task completion results under the two settings. The highest average task success rate is 13.7\% in the Vision-Only setting and 50.1\% with TrajMAS. These results indicate that reliable execution in cloud console GUIs remains challenging for the evaluated models.}

\textbf{The Vision-Only Challenge.} In the Vision-Only setting, even the most capable model (GPT-5.2) achieves only \textbf{13.7\%} average success rate across platforms, with other models performing significantly worse (2.4\%--7.3\%). This stark finding reveals a fundamental limitation: \textit{pure vision-based GUI automation in complex enterprise consoles remains an unsolved challenge}. The difficulty stems from the need to perform simultaneous vision understanding, reasoning, and pixel-level grounding without structural guidance. 

\textbf{Dramatic Performance Gains in TrajMAS Framework.} The introduction of our TrajMAS framework yields transformative improvements. GPT-5.2 achieves the highest average SR of 50.1\%, marking a 3.66$\times$ improvement over its vision-only baseline. The most striking gain is seen in Claude-Sonnet-4.5, which jumps from a nearly non-functional 2.4\% SR to 48.9\% SR (a 20.4$\times$ increase). This indicates that although Claude-Sonnet-4.5 has limited visual grounding ability, providing DOM-based element coordinates effectively unlocks its strong logical reasoning capabilities, allowing it to achieve the second-highest overall average success rate and the best trajectory-quality F1 scores on Alibaba Cloud and Volcengine. Similar dramatic gains are observed across all models: Gemini-3-Flash from 7.3\% to 46.9\% ($6.4\times$) and open-source models like Qwen-3-VL improve from 3.7\% to 26.9\% ($7.3\times$). These results strongly validate our framework design: by providing DOM-based Set-of-Mark grounding, TrajMAS effectively bridges the gap between vision understanding and precise action execution.

\textbf{Platform-Specific Difficulty Variations.} We observe performance variance across cloud platforms in the TrajMAS setting. \textbf{Volcengine} is the most agent-friendly platform, with models achieving the highest SRs (e.g., Claude-Sonnet-4.5: 69.2\%, Seed-1.6: 63.6\%), likely due to its simpler and more consistent UI design. \rev{\textbf{Alibaba Cloud} shows the widest performance spread, from 7.8\% for InternVL to 67.9\% for GPT-5.2.} \textbf{AWS} demonstrates balanced difficulty with top models achieving 34.4\%--56.5\% SR.
\rev{\textbf{Google Cloud} remains challenging, with TrajMAS success rates ranging from 11.8\% to 37.0\%. Dynamic content loading may contribute to this difficulty.}

\section{Threats to Validity}

\textbf{Internal Validity.}
Parts of the annotation pipeline rely on LLM-generated labels. We mitigate this through multi-model consensus, dual-stage verification, and a 20\% human expert review (96\% agreement), though residual label noise may remain. VLM outputs are also non-deterministic; we use greedy decoding (temperature\,=\,0) for reproducibility, but minor stochastic variations across API calls persist.

\textbf{External Validity.}
For Ops-Task Completion, live-console evaluation introduces environmental non-determinism (e.g., variable loading times, session pop-ups) that may affect individual runs, which we mitigate by averaging across multiple tasks per platform and applying consistent timeout policies.

\textbf{Construct Validity.}
For Ops-Action Grounding, our Box-Hit Accuracy is stricter than common distance-based metrics, so alternative metrics could yield different rankings. For Ops-UI Understanding, keyword-matching F1 may not fully capture semantic comprehension, which we partially address by also reporting embedding-based semantic similarity.

\section{Conclusion}

\rev{We introduced GUIOpsBench to evaluate multimodal agents on routine cloud console workflows. The benchmark contains 21,680 samples from four cloud platforms and tests interface understanding, action grounding, and multi-step task completion. Its dual-path collection engine combines SOP-guided trajectories with exploration-driven data generation. Across seven evaluated models, the best average grounding accuracy is 19.09\%, while the highest average task success rate is 13.7\% with vision-only interaction and 50.1\% with TrajMAS. These results highlight the difficulty of connecting operational knowledge with precise GUI interaction and provide a basis for studying more reliable cloud-console agents.}

\section*{\rev{Dataset Availability}}
\rev{The GUIOpsBench dataset is available at \url{https://drive.google.com/drive/folders/158Mzb67Oxt52x2w2OVGVNCv7dRYuxXMu}.}

\bibliographystyle{IEEEtran}
\bibliography{sample-base}

\end{document}

%% file: motivation.tex
{

\section{Motivation}
\label{sec:motivation}

Cloud operations span two broad classes of work: \textit{fault-driven} handling and \textit{routine} operational tasks. Prior AIOps research has concentrated on the former. A systematic mapping of 1{,}079 AIOps studies reports that 62.1\% target failure management (i.e., anomaly detection, root-cause analysis, and failure prediction)~\cite{notaro2021systematic}, and recent benchmarks such as AIOpsLab~\cite{chen2024aiopslab}, OpenRCA~\cite{xu2025openrca}, and OpsEval~\cite{DBLP:conf/sigsoft/LiuPXCSZSZW0LXW25} follow the same emphasis. Routine operational tasks are comparatively under-explored, even though they take up a large share of day-to-day effort. They also differ in a structural way, in that they are carried out through the graphical cloud console rather than through telemetry or APIs. To support these claims with evidence rather than intuition, we conduct an empirical study at Company~$\mathcal{A}$, a large commercial cloud provider. 

\subsection{GUI-centric Routine Operations}
\label{subsec:f1}

We use \textit{GUI-centric routine operation} to mean a recurring, procedural task that an operator completes inside a cloud console by observing page state, reasoning against a Standard Operating Procedure (SOP), and issuing UI actions across one or more pages (Fig.~\ref{fig:background}). Typical instances include provisioning an instance, adjusting a network rule, and configuring a monitoring alert. Unlike a fault-driven incident, such a task has a known goal and a documented procedure.
We illustrate this with a representative on-call case from Company~$\mathcal{A}$.

\textbf{An industrial case.}
\textbf{(1)} On March 9, an on-call SRE received a ticket to bring a new order-query service $S$ online, which required provisioning a compute instance and placing it behind the service load balancer. The SRE opened the console and created an ECS instance, choosing the VPC and vSwitch that already hosted $S$'s peers, binding the \texttt{sg-order} security group, and confirming a pay-as-you-go billing mode. Each choice was made by reading the current page (e.g., which vSwitch already carried $S$'s traffic, whether the security group exposed the service port).
\textbf{(2)} The SRE then added the instance to the backend group of the $S$ load balancer and opened the monitoring page to verify the result. The health-check panel first showed the new instance as unhealthy (red), and the connection heatmap showed no traffic reaching it. Reading these visual signals, the SRE noticed that the security group had not opened the health-check port, corrected it, and watched the indicator turn green and the heatmap fill in.
\textbf{(3)} Only after this visual confirmation did the SRE close the ticket. The whole task (i.e., create, configure, attach, and verify) spanned five console pages and about 15 minutes, and no step was performed through a single scripted API call.

This case is routine rather than exceptional. It carries no failure to diagnose, it follows a documented procedure, and it is still completed end to end in the console. Two properties stand out and motivate the rest of this section. The task directly changes a production system, and its verification depends on reading visual dashboards. In short, a routine operation is a recurring, SOP-guided procedure that spans multiple pages and ends in visual verification rather than one self-contained API call, so evaluating it calls for an environment capturing multi-page navigation and visual verification that telemetry- or API-based benchmarks do not provide.

\subsection{Why Routine Operations Are GUI-centric, Not API-driven}
\label{subsec:f2}

\textbf{Empirical evidence.}
We analyzed recorded operation tickets from Company~$\mathcal{A}$'s production environment over two months. The tickets covered operational changes (37.3\%), configuration changes (32.3\%), version releases (23.8\%), and data changes (3.4\%). Of these, \textbf{89\%} were completed entirely through the GUI, and only 11\% used a CLI, an API, or a mixed workflow. GUI-centric operation is therefore the dominant form of routine cloud work at Company~$\mathcal{A}$, not a stylistic preference. The case in Sec.~\ref{subsec:f1} points to why. Part of the reason is a supply-side gap, and part of it is intrinsic to the tasks.

First, \textbf{many console operations simply have no programmatic counterpart}. Cloud platforms keep evolving their feature sets, and GUI capabilities frequently precede their API or CLI equivalents. AWS, for example, maintains a public CloudFormation Coverage Roadmap that tracks the persistent gap between console-available features and their API counterparts, and acknowledges that full programmatic coverage will take a long time~\cite{awscfncoverage}. Many advanced options, visual editors, and recently launched services remain accessible only through the web console~\cite{avrukin2015sreuis}.

Even where an API exists, two properties of routine tasks make the console the natural medium. \textbf{First, routine tasks change production, so execution must stay observable and interruptible.} They create, delete, or reconfigure live resources (e.g., the instance and security group in Sec.~\ref{subsec:f1}), so a wrong step has a direct blast radius. Consoles keep a human in the loop through confirmation dialogs, diff and impact views, and staged wizards that expose intermediate state before a change is committed~\cite{catchpoint2025srereport}, whereas a raw API call collapses this into a single opaque request that is harder to monitor, pause, or take over mid-task. \textbf{Second, interaction depends on visual dashboards.} Routine tasks usually end by inspecting a dashboard to confirm the intended effect (e.g., the health-check panel and connection heatmap in Sec.~\ref{subsec:f1}). Such dashboards render topology graphs, heatmaps, time-series panels, and color-coded status indicators that compress state otherwise requiring many separate API reads and manual correlation, and they are designed for rapid visual inspection, a mode of reading that text-based output does not support well~\cite{franconeri2021science,few2006information}.

These observations are not specific to Company~$\mathcal{A}$. Voronkov et al.~\cite{voronkov2019firewall} surveyed 300 system administrators and found that 60\% think routine operations should be performed via GUIs, even among self-described command-line users. The Catchpoint SRE Report 2025 likewise notes that SRE teams spend a considerable part of their operational time working in web-based dashboards and consoles~\cite{catchpoint2025srereport}. In short, for routine operations GUI competence is a task requirement rather than an optional convenience: an agent must perceive and act on the console directly, not merely call APIs.

\subsection{Automatable in Demand, Yet Untested}
\label{subsec:f3}

\textbf{Routine operations are worth automating.}
Beyond being GUI-centric, routine tasks are frequent, repetitive, and time-consuming, which makes them attractive automation targets. The Google SRE handbook reports that SREs spend roughly 33\% of their time on \textit{toil}, the manual and repetitive operational work that includes provisioning resources, adjusting configurations, and verifying system states~\cite{beyer2016sre}. Routine tasks also fit automation well, in that they tolerate errors better than incident response (a misconfigured alert can be corrected, a botched incident response may cascade), they follow well-defined SOPs with clear verification criteria, and they recur often enough to amortize the effort. Our interviews echo this priority. Of the 20 SREs, 45\% said automating routine GUI tasks is more necessary than automating fault-driven cases, another 45\% considered the two equally necessary, and only 10\% put fault-driven cases first.
They explained that routine GUI tasks in operations are more repetitive and exhibit more repetitive logic, making them more suitable for automation.
In short, 90\% rated automating routine GUI tasks as at least as necessary as automating fault-driven ones, which agrees with the broader appetite to automate operational toil~\cite{catchpoint2025srereport}.

\textbf{Yet cloud consoles differ from the GUIs current benchmarks cover.}
Recent advances in vision-language models (VLMs) make console automation increasingly plausible, but existing GUI-agent benchmarks are built around generic web, mobile, and desktop applications~\cite{deng2023mind2webgeneralistagentweb,koh2024visualwebarenaevaluatingmultimodalagents,yao2023webshopscalablerealworldweb,rawles2023androidwildlargescaledataset,mu2025gui360circcomprehensivedatasetbenchmark,xie2024osworld}. Cloud consoles differ from these along three axes, so results on general GUIs do not transfer and a separate benchmark is needed. \textbf{First, dense domain-specific visuals with graphs and charts:} a large part of a console's information appears as topology graphs, monitoring charts, time series, heatmaps, and configuration panels (e.g., Fig.~\ref{fig:tasks} and the dashboards in Sec.~\ref{subsec:f1}) that must be read and acted on, yet are uncommon in general GUI benchmarks built on consumer apps and sites~\cite{mu2025gui360circcomprehensivedatasetbenchmark,xie2024osworld}. \textbf{Second, SOP-driven workflows with domain knowledge:} a console task is usually tied to a ticket and its SOP, and completing it requires combining that SOP with domain knowledge to plan and verify a sequence of steps, whereas general GUI tasks take a single self-contained instruction (e.g., ``click the buy button'') and depend mostly on the current page. \textbf{Third, deep resource dependency:} a single operation often touches several coupled cloud resources (e.g., ECS, VPC, security group, IAM), so acting correctly requires reasoning about these dependencies rather than treating each page in isolation.

Existing AIOps benchmarks do not close this gap either. AIOpsLab~\cite{chen2024aiopslab}, ITBench~\cite{jha2025itbenchevaluatingaiagents}, and OpsEval~\cite{DBLP:conf/sigsoft/LiuPXCSZSZW0LXW25} evaluate agents on text-centric, API-driven tasks and do not test GUI interaction at all. As a result, the ability of modern VLMs to carry out routine GUI-centric cloud operations remains untested. This motivates a vision-centric benchmark tailored to cloud consoles that tests whether agents can execute and verify real Ops procedures through GUIs, a gap \benchname{} is designed to fill.

}

%% file: tab/grounding.tex
\begin{table}[t]
    \centering
    \caption{Grounding accuracy across different cloud platforms. }
    \label{tab:grounding_results}
    \resizebox{\columnwidth}{!}{%
    \begin{tabular}{lccccc}
    \toprule
    \textbf{Model} & \textbf{Alibaba} & \textbf{AWS} & \textbf{Google} & \textbf{VolcEngine} & \textbf{Average} \\
    \midrule
    GPT-5.2 & \textbf{21.60} & \textbf{18.07} & \textbf{18.46} & \textbf{18.68} & \textbf{19.09} \\
    Claude-Sonnet-4.5  & 1.18 & 0.62 & 2.28 & 0.18 & 1.07 \\
    Gemini-3-Flash   & 1.48 & 0.62 & 8.09 & 7.00 & 4.27 \\
    GPT-4o  & 5.88 & 1.91 & 8.64 & 10.54 & 7.69 \\
    Seed-1.6   & 0.59 & 0.62 & 3.73 & 0.92 & 1.60 \\
    InternVL    & 0.59 & 1.87 & 2.49 & 0.00 & 1.19 \\
    Qwen-3-VL  & 1.18 & 0.31 & 5.60 & 0.73 & 2.13 \\
    \bottomrule
    \end{tabular}%
    }
    
    \footnotesize
\end{table}

%% file: tab/task-completion.tex
\definecolor{visionbg}{RGB}{235, 245, 255} 
\definecolor{trajbg}{RGB}{235, 255, 235}   

\begin{table*}[t]
\centering
\caption{Performance evaluation on different platforms. We report Success Rate (SR), Precision (Pr), Recall (Rc), and F1-score (F1). }
\label{tab:task_completion}
\resizebox{\textwidth}{!}{%
\begin{tabular}{l cccc cccc cccc cccc c}
\toprule
\multirow{2}{*}{\textbf{Model}} & \multicolumn{4}{c}{\textbf{Alibaba Cloud}} & \multicolumn{4}{c}{\textbf{AWS}} & \multicolumn{4}{c}{\textbf{Google Cloud}} & \multicolumn{4}{c}{\textbf{Volcengine}} & \multirow{2}{*}{\textbf{Final Avg}} \\

\cmidrule(lr){2-5} \cmidrule(lr){6-9} \cmidrule(lr){10-13} \cmidrule(lr){14-17}

 & SR & Pr & Rc & F1 & SR & Pr & Rc & F1 & SR & Pr & Rc & F1 & SR & Pr & Rc & F1 & (SR) \\
\midrule

\rowcolor{visionbg} \multicolumn{18}{c}{\textbf{\textit{VisionOnly Setting}}} \\
GPT-5.2 & \textbf{\underline{13.8}} & \textbf{\underline{13.7}} & \textbf{\underline{14.6}} & \textbf{\underline{14.1}} & \textbf{\underline{14.3}} & 10.2 & 10.8 & 10.4 & \textbf{\underline{12.5}} & \textbf{\underline{12.9}} & \textbf{\underline{13.0}} & \textbf{\underline{12.9}} & \textbf{\underline{14.0}} & \textbf{\underline{13.5}} & \textbf{\underline{14.0}} & \textbf{\underline{13.6}} & \textbf{\underline{13.7}} \\
Claude-Sonnet-4.5 & 1.4 & 6.1 & 7.5 & 6.7 & 2.9 & 6.3 & 7.6 & 6.9 & 1.8 & 6.6 & 7.4 & 6.9 & 3.3 & 7.1 & 8.6 & 7.8 & 2.4 \\
Gemini-3-Flash & 5.9 & 8.4 & 9.5 & 8.9 & 10.3 & \textbf{\underline{12.1}} & \textbf{\underline{13.9}} & \textbf{\underline{12.9}} & 5.1 & 7.6 & 8.6 & 8.0 & 7.9 & 10.7 & 12.0 & 11.3 & 7.3 \\
GPT-4o & 6.5 & 7.5 & 9.0 & 8.1 & 3.0 & 7.3 & 8.7 & 7.9 & 5.8 & 8.1 & 9.2 & 8.6 & 11.1 & 11.3 & 12.1 & 11.6 & 6.6 \\
SEED-1.6 & 1.5 & 6.2 & 7.4 & 6.7 & 5.8 & 8.1 & 8.7 & 8.3 & 3.2 & 6.7 & 7.8 & 7.1 & 10.8 & 11.0 & 12.0 & 11.4 & 5.3 \\
InternVL & 1.4 & 6.1 & 7.5 & 6.7 & 5.6 & 7.5 & 8.3 & 7.8 & 1.6 & 6.0 & 7.5 & 6.6 & 6.3 & 8.0 & 9.2 & 8.5 & 3.7 \\
Qwen-3-VL & 1.7 & 6.6 & 7.9 & 7.2 & 2.2 & 5.9 & 7.3 & 6.5 & 1.9 & 6.6 & 8.1 & 7.2 & 8.9 & 10.0 & 11.3 & 10.6 & 3.7 \\

\midrule

\rowcolor{trajbg} \multicolumn{18}{c}{\textbf{\textit{TrajMAS Setting}}} \\
GPT-5.2 & \textbf{\underline{67.9}} & 12.8 & 12.6 & 12.7 & 50.0 & 16.5 & 19.8 & 16.9 & 26.3 & 17.8 & 19.1 & 18.4 & 56.0 & 37.0 & 36.7 & 35.1 & \textbf{\underline{50.1}} \\

Claude-Sonnet-4.5 & 54.8 & \textbf{\underline{19.4}} & 23.4 & \textbf{\underline{19.1}} & 34.4 & \textbf{\underline{23.2}} & 20.9 & 20.7 & \textbf{\underline{37.0}} & 23.5 & 29.3 & 24.1 & \textbf{\underline{69.2}} & \textbf{\underline{47.0}} & \textbf{\underline{44.4}} & \textbf{\underline{44.5}} & 48.9 \\
Gemini-3-Flash & 46.2 & 13.9 & 24.4 & 15.1 & \textbf{\underline{56.5}} & 19.4 & \textbf{\underline{24.8}} & 21.1 & 31.5 & 22.2 & 26.1 & 22.4 & 53.3 & 34.0 & 36.0 & 33.4 & 46.9 \\
GPT-4o & 19.6 & 8.9 & 10.0 & 9.2 & 41.9 & 15.9 & 23.4 & 18.0 & 21.1 & 23.5 & \textbf{\underline{35.7}} & \textbf{\underline{26.3}} & 31.4 & 31.7 & 39.6 & 33.1 & 28.5 \\
SEED-1.6 & 11.4 & 7.9 & 11.8 & 7.6 & 22.6 & 8.2 & 8.8 & 7.8 & 27.8 & \textbf{\underline{26.1}} & 26.9 & 24.8 & 63.6 & 39.7 & 40.9 & 39.4 & 31.4 \\

InternVL & 7.8 & 10.9 & 10.5 & 9.6 & 31.8 & 15.2 & 22.6 & 16.8 & 11.8 & 19.0 & 27.3 & 20.8 & 48.5 & 34.3 & 39.1 & 35.4 & 25.0 \\
Qwen-3-VL & 13.3 & 10.1 & 10.6 & 9.8 & 21.7 & 22.1 & 22.9 & \textbf{\underline{21.5}} & 18.9 & 25.3 & 30.2 & 25.3 & 53.7 & 35.4 & 34.2 & 33.8 & 26.9 \\

\bottomrule
\end{tabular}%
}
\end{table*}

%% file: sample-base.bib
@inproceedings{logad,
author = {Yu, Boxi and Yao, Jiayi and Fu, Qiuai and Zhong, Zhiqing and Xie, Haotian and Wu, Yaoliang and Ma, Yuchi and He, Pinjia},
title = {Deep Learning or Classical Machine Learning? An Empirical Study on Log-Based Anomaly Detection},
year = {2024},
isbn = {9798400702174},
publisher = {Association for Computing Machinery},
address = {New York, NY, USA},
url = {https://doi.org/10.1145/3597503.3623308},
doi = {10.1145/3597503.3623308},
booktitle = {Proceedings of the IEEE/ACM 46th International Conference on Software Engineering},
articleno = {35},
numpages = {13},
location = {Lisbon, Portugal},
series = {ICSE '24}
}

@inproceedings{kon2024iaceval,
  title     = {{IaC-Eval}: A Code Generation Benchmark for Cloud Infrastructure-as-Code Programs},
  author    = {Patrick Tser Jern Kon and Jiachen Liu and Yiming Qiu and Weijun Fan and Ting He and Lei Lin and Haoran Zhang and Owen M. Park and George Sajan Elengikal and Yuxin Kang and Ang Chen and Mosharaf Chowdhury and Myungjin Lee and Xinyu Wang},
  booktitle = {Advances in Neural Information Processing Systems 37 (NeurIPS 2024), Datasets and Benchmarks Track},
  year      = {2024},
  url       = {https://github.com/autoiac-project/iac-eval}
}

@misc{yang2023setofmarkpromptingunleashesextraordinary,
      title={Set-of-Mark Prompting Unleashes Extraordinary Visual Grounding in GPT-4V}, 
      author={Jianwei Yang and Hao Zhang and Feng Li and Xueyan Zou and Chunyuan Li and Jianfeng Gao},
      year={2023},
      eprint={2310.11441},
      archivePrefix={arXiv},
      primaryClass={cs.CV},
      url={https://arxiv.org/abs/2310.11441}, 
}

@inproceedings{voronkov2019firewall,
title = {System Administrators Prefer Command Line Interfaces, Don't They? An Exploratory Study of Firewall Interfaces},
author = {Artem Voronkov and Leonardo A. Martucci and Stefan Lindskog},
booktitle = {Fifteenth Symposium on Usable Privacy and Security (SOUPS 2019)},
year = {2019},
pages = {259--271},
isbn = {978-1-939133-05-2},
address = {Santa Clara, CA},
publisher = {USENIX Association},
month = aug,
url = {https://www.usenix.org/conference/soups2019/presentation/voronkov}
}

@inproceedings{avrukin2015sreuis,
title = {{SRE UIs---Transcending the CLI}},
author = {Michael Avrukin},
booktitle = {USENIX SREcon Europe 2015},
year = {2015},
address = {Dublin, Ireland},
publisher = {USENIX Association},
month = may,
url = {https://www.usenix.org/conference/srecon15europe/program/presentation/avrukin}
}

@misc{catchpoint2025srereport,
title = {The SRE Report 2025},
author = {Catchpoint},
year = {2025},
howpublished = {Catchpoint Systems Report},
url = {https://resources.catchpoint.com/hubfs/Website%20Assets%20-%20Briefs%2C%20EBooks%2C%20etc/The%20SRE%20Report%202025%20Catchpoint.pdf}
}

@misc{gpt4o,
  author={OpenAI},
  title={Hello GPT-4o},
  url={https://openai.com/index/hello-gpt-4o/},
  year=2024,
}

@misc{zhang2024surveyaiopsfailuremanagement,
      title={A Survey of AIOps for Failure Management in the Era of Large Language Models}, 
      author={Lingzhe Zhang and Tong Jia and Mengxi Jia and Yifan Wu and Aiwei Liu and Yong Yang and Zhonghai Wu and Xuming Hu and Philip S. Yu and Ying Li},
      year={2024},
      eprint={2406.11213},
      archivePrefix={arXiv},
      primaryClass={cs.SE},
      url={https://arxiv.org/abs/2406.11213}, 
}

@article{ma2020diagnosing,
  title={Diagnosing root causes of intermittent slow queries in cloud databases},
  author={Ma, Minghua and Yin, Zheng and Zhang, Shenglin and Wang, Sheng and Zheng, Christopher and Jiang, Xinhao and Hu, Hanwen and Luo, Cheng and Li, Yilin and Qiu, Nengjun and others},
  journal={Proceedings of the VLDB Endowment},
  volume={13},
  number={8},
  pages={1176--1189},
  year={2020},
  publisher={VLDB Endowment}
}

@misc{claude-4.5,
  author={Anthropic},
  title={Introducing Claude Sonnet 4.5},
  url={https://www.anthropic.com/news/claude-sonnet-4-5},
  year=2025,
}

@misc{llama4,
  author={Meta AI},
  title={The Llama 4 herd: The beginning of a new era of natively multimodal AI innovation},
  url={https://ai.meta.com/blog/llama-4-multimodal-intelligence/},
  year=2025,
}

@misc{bai2025qwen25vltechnicalreport,
      title={Qwen2.5-VL Technical Report}, 
      author={Shuai Bai and Keqin Chen and Xuejing Liu and Jialin Wang and Wenbin Ge and Sibo Song and Kai Dang and Peng Wang and Shijie Wang and Jun Tang and Humen Zhong and Yuanzhi Zhu and Mingkun Yang and Zhaohai Li and Jianqiang Wan and Pengfei Wang and Wei Ding and Zheren Fu and Yiheng Xu and Jiabo Ye and Xi Zhang and Tianbao Xie and Zesen Cheng and Hang Zhang and Zhibo Yang and Haiyang Xu and Junyang Lin},
      year={2025},
      eprint={2502.13923},
      archivePrefix={arXiv},
      primaryClass={cs.CV},
      url={https://arxiv.org/abs/2502.13923}, 
}

@misc{chen2024internvlscalingvisionfoundation,
      title={InternVL: Scaling up Vision Foundation Models and Aligning for Generic Visual-Linguistic Tasks}, 
      author={Zhe Chen and Jiannan Wu and Wenhai Wang and Weijie Su and Guo Chen and Sen Xing and Muyan Zhong and Qinglong Zhang and Xizhou Zhu and Lewei Lu and Bin Li and Ping Luo and Tong Lu and Yu Qiao and Jifeng Dai},
      year={2024},
      eprint={2312.14238},
      archivePrefix={arXiv},
      primaryClass={cs.CV},
      url={https://arxiv.org/abs/2312.14238}, 
}

@misc{bai2025qwen3vltechnicalreport,
      title={Qwen3-VL Technical Report}, 
      author={Shuai Bai and Yuxuan Cai and Ruizhe Chen and Keqin Chen and Xionghui Chen and Zesen Cheng and Lianghao Deng and Wei Ding and Chang Gao and Chunjiang Ge and Wenbin Ge and Zhifang Guo and Qidong Huang and Jie Huang and Fei Huang and Binyuan Hui and Shutong Jiang and Zhaohai Li and Mingsheng Li and Mei Li and Kaixin Li and Zicheng Lin and Junyang Lin and Xuejing Liu and Jiawei Liu and Chenglong Liu and Yang Liu and Dayiheng Liu and Shixuan Liu and Dunjie Lu and Ruilin Luo and Chenxu Lv and Rui Men and Lingchen Meng and Xuancheng Ren and Xingzhang Ren and Sibo Song and Yuchong Sun and Jun Tang and Jianhong Tu and Jianqiang Wan and Peng Wang and Pengfei Wang and Qiuyue Wang and Yuxuan Wang and Tianbao Xie and Yiheng Xu and Haiyang Xu and Jin Xu and Zhibo Yang and Mingkun Yang and Jianxin Yang and An Yang and Bowen Yu and Fei Zhang and Hang Zhang and Xi Zhang and Bo Zheng and Humen Zhong and Jingren Zhou and Fan Zhou and Jing Zhou and Yuanzhi Zhu and Ke Zhu},
      year={2025},
      eprint={2511.21631},
      archivePrefix={arXiv},
      primaryClass={cs.CV},
      url={https://arxiv.org/abs/2511.21631}, 
}

@misc{wang2024officebenchbenchmarkinglanguageagents,
      title={OfficeBench: Benchmarking Language Agents across Multiple Applications for Office Automation}, 
      author={Zilong Wang and Yuedong Cui and Li Zhong and Zimin Zhang and Da Yin and Bill Yuchen Lin and Jingbo Shang},
      year={2024},
      eprint={2407.19056},
      archivePrefix={arXiv},
      primaryClass={cs.CL},
      url={https://arxiv.org/abs/2407.19056}, 
}

@misc{koh2024visualwebarenaevaluatingmultimodalagents,
      title={VisualWebArena: Evaluating Multimodal Agents on Realistic Visual Web Tasks}, 
      author={Jing Yu Koh and Robert Lo and Lawrence Jang and Vikram Duvvur and Ming Chong Lim and Po-Yu Huang and Graham Neubig and Shuyan Zhou and Ruslan Salakhutdinov and Daniel Fried},
      year={2024},
      eprint={2401.13649},
      archivePrefix={arXiv},
      primaryClass={cs.LG},
      url={https://arxiv.org/abs/2401.13649}, 
}

@misc{yao2023webshopscalablerealworldweb,
      title={WebShop: Towards Scalable Real-World Web Interaction with Grounded Language Agents}, 
      author={Shunyu Yao and Howard Chen and John Yang and Karthik Narasimhan},
      year={2023},
      eprint={2207.01206},
      archivePrefix={arXiv},
      primaryClass={cs.CL},
      url={https://arxiv.org/abs/2207.01206}, 
}

@misc{deng2023mind2webgeneralistagentweb,
      title={Mind2Web: Towards a Generalist Agent for the Web}, 
      author={Xiang Deng and Yu Gu and Boyuan Zheng and Shijie Chen and Samuel Stevens and Boshi Wang and Huan Sun and Yu Su},
      year={2023},
      eprint={2306.06070},
      archivePrefix={arXiv},
      primaryClass={cs.CL},
      url={https://arxiv.org/abs/2306.06070}, 
}

@misc{mu2025gui360circcomprehensivedatasetbenchmark,
      title={GUI-360$^\circ$: A Comprehensive Dataset and Benchmark for Computer-Using Agents}, 
      author={Jian Mu and Chaoyun Zhang and Chiming Ni and Lu Wang and Bo Qiao and Kartik Mathur and Qianhui Wu and Yuhang Xie and Xiaojun Ma and Mengyu Zhou and Si Qin and Liqun Li and Yu Kang and Minghua Ma and Qingwei Lin and Saravan Rajmohan and Dongmei Zhang},
      year={2025},
      eprint={2511.04307},
      archivePrefix={arXiv},
      primaryClass={cs.AI},
      url={https://arxiv.org/abs/2511.04307}, 
}

@misc{zhang2024androidzoochainofactionthoughtgui,
      title={Android in the Zoo: Chain-of-Action-Thought for GUI Agents}, 
      author={Jiwen Zhang and Jihao Wu and Yihua Teng and Minghui Liao and Nuo Xu and Xiao Xiao and Zhongyu Wei and Duyu Tang},
      year={2024},
      eprint={2403.02713},
      archivePrefix={arXiv},
      primaryClass={cs.CL},
      url={https://arxiv.org/abs/2403.02713}, 
}

@misc{rawles2023androidwildlargescaledataset,
      title={Android in the Wild: A Large-Scale Dataset for Android Device Control}, 
      author={Christopher Rawles and Alice Li and Daniel Rodriguez and Oriana Riva and Timothy Lillicrap},
      year={2023},
      eprint={2307.10088},
      archivePrefix={arXiv},
      primaryClass={cs.LG},
      url={https://arxiv.org/abs/2307.10088}, 
}

@inproceedings{10.1145/3126594.3126651,
author = {Deka, Biplab and Huang, Zifeng and Franzen, Chad and Hibschman, Joshua and Afergan, Daniel and Li, Yang and Nichols, Jeffrey and Kumar, Ranjitha},
title = {Rico: A Mobile App Dataset for Building Data-Driven Design Applications},
year = {2017},
isbn = {9781450349819},
publisher = {Association for Computing Machinery},
address = {New York, NY, USA},
url = {https://doi.org/10.1145/3126594.3126651},
doi = {10.1145/3126594.3126651},
booktitle = {Proceedings of the 30th Annual ACM Symposium on User Interface Software and Technology},
pages = {845--854},
numpages = {10},
location = {Qu\'{e}bec City, QC, Canada},
series = {UIST '17}
}

@inproceedings{chen2024aiopslab,
  title = {AIOpsLab: A Holistic Framework to Evaluate AI Agents for Enabling Autonomous Clouds},
  author = {Chen, Yinfang and Shetty, Manish and Somashekar, Gagan and Ma, Minghua and Simmhan, Yogesh and Mace, Jonathan and Bansal, Chetan and Wang, Rujia and Rajmohan, Saravan},
  year = {2025},
  booktitle = {Proceedings of the Eighth Annual Conference on Machine Learning and Systems (MLSys'25)},
  month = may,
  github = {https://github.com/microsoft/AIOpsLab},
  url = {https://www.microsoft.com/en-us/research/publication/aiopslab-a-holistic-framework-for-evaluating-ai-agents-for-enabling-autonomous-cloud/},
}

@inproceedings{
jha2025itbenchevaluatingaiagents,
title={{ITB}ench: Evaluating {AI} Agents across Diverse Real-World {IT} Automation Tasks},
author={Saurabh Jha and Rohan R. Arora and Yuji Watanabe and Takumi Yanagawa and Yinfang Chen and Jackson Clark and Bhavya Bhavya and Mudit Verma and Harshit Kumar and Hirokuni Kitahara and Noah Zheutlin and Saki Takano and Divya Pathak and Felix George and Xinbo Wu and Bekir O Turkkan and Gerard Vanloo and Michael Nidd and Ting Dai and Oishik Chatterjee and Pranjal Gupta and Suranjana Samanta and Pooja Aggarwal and Rong Lee and Jae-wook Ahn and Debanjana Kar and Amit Paradkar and Yu Deng and Pratibha Moogi and Prateeti Mohapatra and Naoki Abe and Chandrasekhar Narayanaswami and Tianyin Xu and Lav R. Varshney and Ruchi Mahindru and Anca Sailer and Laura Shwartz and Daby Sow and Nicholas C. M. Fuller and Ruchir Puri},
booktitle={Forty-second International Conference on Machine Learning},
year={2025},
url={https://openreview.net/forum?id=jP59rz1bZk}
}

@misc{pujar2023ansiblewisdom,
      title={Automated Code generation for Information Technology Tasks in YAML through Large Language Models}, 
      author={Saurabh Pujar and Luca Buratti and Xiaojie Guo and Nicolas Dupuis and Burn Lewis and Sahil Suneja and Atin Sood and Ganesh Nalawade and Matthew Jones and Alessandro Morari and Ruchir Puri},
      year={2023},
      eprint={2305.02783},
      archivePrefix={arXiv},
      primaryClass={cs.SE},
      url={https://arxiv.org/abs/2305.02783}, 
}

@inproceedings{chen2023automatic,
  title = {Automatic Root Cause Analysis via Large Language Models for Cloud Incidents},
  author = {Chen, Yinfang and Xie, Huaibing and Ma, Minghua and Kang, Yu and Gao, Xin and Shi, Liu and Cao, Yunjie and Gao, Xuedong and Fan, Hao and Wen, Ming and Zeng, Jun and Ghosh, Supriyo and Zhang, Xuchao and Zhang, Chaoyun and Lin, Qingwei and Rajmohan, Saravan and Zhang, Dongmei and Xu, Tianyin},
  booktitle = {Proceedings of the 19th European Conference on Computer Systems (EuroSys'24)},
  year = {2024},
  month = apr,
}

@inproceedings{DBLP:conf/sigsoft/LiuPXCSZSZW0LXW25,
  author       = {Yuhe Liu and
                  Changhua Pei and
                  Longlong Xu and
                  Bohan Chen and
                  Mingze Sun and
                  Zhirui Zhang and
                  Yongqian Sun and
                  Shenglin Zhang and
                  Kun Wang and
                  Haiming Zhang and
                  Jianhui Li and
                  Gaogang Xie and
                  Xidao Wen and
                  Xiaohui Nie and
                  Minghua Ma and
                  Dan Pei},
  editor       = {Leonardo Montecchi and
                  Jingyue Li and
                  Denys Poshyvanyk and
                  Dongmei Zhang},
  title        = {OpsEval: {A} Comprehensive Benchmark Suite for Evaluating Large Language
                  Models' Capability in {IT} Operations Domain},
  booktitle    = {Proceedings of the 33rd {ACM} International Conference on the Foundations
                  of Software Engineering, {FSE} Companion 2025, Clarion Hotel Trondheim,
                  Trondheim, Norway, June 23-28, 2025},
  pages        = {503--513},
  publisher    = {{ACM}},
  year         = {2025},
  url          = {https://doi.org/10.1145/3696630.3728572},
  doi          = {10.1145/3696630.3728572},
  bibsource    = {dblp computer science bibliography, https://dblp.org}
}

@misc{flowofaction,
      title={Flow-of-Action: SOP Enhanced LLM-Based Multi-Agent System for Root Cause Analysis}, 
      author={Changhua Pei and Zexin Wang and Fengrui Liu and Zeyan Li and Yang Liu and Xiao He and Rong Kang and Tieying Zhang and Jianjun Chen and Jianhui Li and Gaogang Xie and Dan Pei},
      year={2025},
      eprint={2502.08224},
      archivePrefix={arXiv},
      primaryClass={cs.SE},
      url={https://arxiv.org/abs/2502.08224}, 
}

@misc{rcacopilot,
      title={Automatic Root Cause Analysis via Large Language Models for Cloud Incidents}, 
      author={Yinfang Chen and Huaibing Xie and Minghua Ma and Yu Kang and Xin Gao and Liu Shi and Yunjie Cao and Xuedong Gao and Hao Fan and Ming Wen and Jun Zeng and Supriyo Ghosh and Xuchao Zhang and Chaoyun Zhang and Qingwei Lin and Saravan Rajmohan and Dongmei Zhang and Tianyin Xu},
      year={2023},
      eprint={2305.15778},
      archivePrefix={arXiv},
      primaryClass={cs.SE},
      url={https://arxiv.org/abs/2305.15778}, 
}

@misc{mabc,
      title={mABC: multi-Agent Blockchain-Inspired Collaboration for root cause analysis in micro-services architecture}, 
      author={Wei Zhang and Hongcheng Guo and Jian Yang and Zhoujin Tian and Yi Zhang and Chaoran Yan and Zhoujun Li and Tongliang Li and Xu Shi and Liangfan Zheng and Bo Zhang},
      year={2024},
      eprint={2404.12135},
      archivePrefix={arXiv},
      primaryClass={cs.MA},
      url={https://arxiv.org/abs/2404.12135}, 
}

@inproceedings{rcagent, series={CIKM '24},
   title={RCAgent: Cloud Root Cause Analysis by Autonomous Agents with Tool-Augmented Large Language Models},
   url={http://dx.doi.org/10.1145/3627673.3680016},
   DOI={10.1145/3627673.3680016},
   booktitle={Proceedings of the 33rd ACM International Conference on Information and Knowledge Management},
   publisher={ACM},
   author={Wang, Zefan and Liu, Zichuan and Zhang, Yingying and Zhong, Aoxiao and Wang, Jihong and Yin, Fengbin and Fan, Lunting and Wu, Lingfei and Wen, Qingsong},
   year={2024},
   month=oct, pages={4966--4974},
   collection={CIKM '24} }

@article{zhou2018fault,
  title={Fault analysis and debugging of microservice systems: Industrial survey, benchmark system, and empirical study},
  author={Zhou, Xiang and Peng, Xin and Xie, Tao and Sun, Jun and Ji, Chao and Li, Wenhai and Ding, Dan},
  journal={IEEE Transactions on Software Engineering},
  volume={47},
  number={2},
  pages={243--260},
  year={2018},
  publisher={IEEE}
}

@misc{doubao16,
    title={Seed1.6},
    author={ByteDance Seed},
    url={https://seed.bytedance.com/en/seed1_6},
    note={Accessed: 2026-01-28}
}

@misc{gpt52,
    title={Introducing GPT-5.2},
    author={OpenAI},
    url={https://openai.com/index/introducing-gpt-5-2/},
    note={Accessed: 2026-01-28}
}

@misc{aws,
    title={AWS},
    author={Amazon Web Services},
    url={https://aws.amazon.com/},
    note={Accessed: 2026-01-28}
}

@misc{gcloud,
    title={Google Cloud},
    author={Google},
    url={https://cloud.google.com/},
    note={Accessed: 2026-01-28}
}

@misc{alicloud,
    title={Alibaba Cloud},
    author={Alibaba},
    url={https://www.alibabacloud.com/},
    note={Accessed: 2026-01-28}
}

@misc{volcengine,
    title={Volcano Engine},
    author={Bytedance},
    url={https://www.volcengine.com/},
    note={Accessed: 2026-01-28}
}

@book{few2006information,
    title={Information Dashboard Design: The Effective Visual Communication of Data},
    author={Few, Stephen},
    year={2006},
    publisher={O'Reilly Media}
}

@misc{awscfncoverage,
    title={The History and Future Roadmap of the {AWS CloudFormation} Registry},
    author={{Amazon Web Services}},
    year={2023},
    howpublished={\url{https://aws.amazon.com/blogs/devops/cloudformation-coverage/}},
    note={Accessed: 2026-03-25}
}

@article{franconeri2021science,
    title={The Science of Visual Data Communication: What Works},
    author={Franconeri, Steven L. and Padilla, Lace M. and Shah, Priti and Zacks, Jeffrey M. and Hullman, Jessica},
    journal={Psychological Science in the Public Interest},
    volume={22},
    number={3},
    pages={110--161},
    year={2021},
    publisher={SAGE Publications},
    doi={10.1177/15291006211051956}
}

@inproceedings{notaro2021systematic,
    title={A Systematic Mapping Study in {AIOps}},
    author={Notaro, Paolo and Cardoso, Jorge and Gerndt, Michael},
    booktitle={ICSOC 2020 Workshops, LNCS 12632},
    pages={110--123},
    year={2021},
    publisher={Springer},
    doi={10.1007/978-3-030-76352-7\_15}
}

@book{beyer2016sre,
    title={Site Reliability Engineering: How {Google} Runs Production Systems},
    author={Beyer, Betsy and Jones, Chris and Petoff, Jennifer and Murphy, Niall Richard},
    year={2016},
    publisher={O'Reilly Media},
    url={https://sre.google/sre-book/eliminating-toil/}
}

@inproceedings{xu2025openrca,
  title={OpenRCA: Can Large Language Models Locate the Root Cause of Software Failures?},
  author={Xu, Junjielong and Zhang, Qinan and Zhong, Zhiqing and He, Shilin and Zhang, Chaoyun and Lin, Qingwei and Pei, Dan and He, Pinjia and Zhang, Dongmei and Zhang, Qi},
  booktitle={The Thirteenth International Conference on Learning Representations},
  year={2025},
  url={https://openreview.net/forum?id=M4qNIzQYpd}
}

@inproceedings{xie2024osworld,
  title={OSWorld: Benchmarking Multimodal Agents for Open-Ended Tasks in Real Computer Environments},
  author={Tianbao Xie and Danyang Zhang and Jixuan Chen and Xiaochuan Li and Siheng Zhao and Ruisheng Cao and Toh Jing Hua and Zhoujun Cheng and Dongchan Shin and Fangyu Lei and Yitao Liu and Yiheng Xu and Shuyan Zhou and Silvio Savarese and Caiming Xiong and Victor Zhong and Tao Yu},
  booktitle={Advances in Neural Information Processing Systems 37 (NeurIPS 2024), Datasets and Benchmarks Track},
  year={2024},
  url={https://arxiv.org/abs/2404.07972}
}

@inproceedings{drouin2024workarena,
  title={WorkArena: How Capable Are Web Agents at Solving Common Knowledge Work Tasks?},
  author={Drouin, Alexandre and Gasse, Maxime and Caccia, Massimo and Laradji, Issam H. and Del Verme, Manuel and Marty, Tom and Boisvert, L\'{e}o and Thakkar, Megh and Cappart, Quentin and Vazquez, David and Chapados, Nicolas and Lacoste, Alexandre},
  booktitle={Advances in Neural Information Processing Systems 37 (NeurIPS 2024), Datasets and Benchmarks Track},
  year={2024},
  url={https://arxiv.org/abs/2403.07718}
}

@misc{ecs,
  title = {Elastic Common Schema: Normalizing your data with ECS.},
  howpublished = {https://www.elastic.co/elasticsearch/common-schema},
  note = {Accessed: 2025/4/22},
  author = {Elasticsearch},
  year = {2025}
}

@inproceedings{nezha,
author = {Yu, Guangba and Chen, Pengfei and Li, Yufeng and Chen, Hongyang and Li, Xiaoyun and Zheng, Zibin},
title = {Nezha: Interpretable Fine-Grained Root Causes Analysis for Microservices on Multi-modal Observability Data},
year = {2023},
isbn = {9798400703270},
publisher = {Association for Computing Machinery},
address = {New York, NY, USA},
url = {https://doi.org/10.1145/3611643.3616249},
doi = {10.1145/3611643.3616249},
booktitle = {Proceedings of the 31st ACM Joint European Software Engineering Conference and Symposium on the Foundations of Software Engineering},
pages = {553--565},
numpages = {13},
location = {San Francisco, CA, USA},
series = {ESEC/FSE 2023}
}
